\documentclass[gmd, manuscript]{copernicus}

\usepackage{array}
\usepackage{booktabs}
\usepackage{listings}
\usepackage{stmaryrd}
\usepackage{graphicx}
\usepackage{mathtools}
\usepackage{listing}
\usepackage{caption}
\usepackage{subcaption}

\DeclarePairedDelimiter\jump{\llbracket}{\rrbracket}

\DeclarePairedDelimiter\cellinner{(}{)}
\DeclarePairedDelimiter{\facetinner}{\langle}{\rangle}
\newcommand{\Hdiv}{\texttt{HDiv}}
\newcommand{\Hcurl}{\texttt{HCurl}}

\newcommand{\zhat}{\hat{\boldsymbol{z}}}

\definecolor{DarkBlue}{rgb}{0.00,0.00,0.55}
\definecolor{DarkRed}{rgb}{0.55,0.00,0.00}
\definecolor{DarkGreen}{rgb}{0.00,0.55,0.00}
\definecolor{Purple}{rgb}{0.5, 0.0, 0.5}
\definecolor{BurntOrange}{RGB}{204,85,0}
\definecolor{backcolour}{rgb}{0.95,0.95,0.92}

\lstdefinelanguage[firedrake]{python}[]{python}{%
	emph=[3]{Tensor,AssembledVector,block,T,inv,decomposition},
	emph=[2]{Function,FunctionSpace,VectorFunctionSpace,Constant,solve,assemble},
	emph={dot,div,grad,inner,Coefficient,FiniteElement,TrialFunction,TestFunction,TrialFunctions,TestFunctions,triangle,jump,dx,dS,ds,FacetNormal,split}
}

\lstdefinestyle{framed}{frame=tb}

\begin{document}
\nolinenumbers
% Title is a work in-progress...
\title{Slate: extending Firedrake's domain-specific abstraction to hybridized solvers for geoscience and beyond}

% \Author[affil]{given_name}{surname}

\Author[1]{T. H.}{Gibson}
\Author[2]{L.}{Mitchell}
\Author[1]{D. A.}{Ham}
\Author[1]{C. J.}{Cotter}

\affil[1]{Department of Mathematics, Imperial College London, London, SW7 2AZ, UK}
\affil[2]{Department of Computer Science, Durham University, Durham, DH1 3LE, UK}

%% The [] brackets identify the author with the corresponding affiliation. 1, 2, 3, etc. should be inserted.

\runningtitle{Hybridization and static condensation methods for GFD}

\runningauthor{Gibson et al}

\correspondence{T. H. Gibson (t.gibson15@imperial.ac.uk)}

\received{}
\pubdiscuss{} %% only important for two-stage journals
\revised{}
\accepted{}
\published{}

%% These dates will be inserted by Copernicus Publications during the typesetting process.

\firstpage{1}

\maketitle

\begin{abstract}
Within the finite element community, discontinuous Galerkin (DG) and mixed finite element methods have become increasingly
popular in simulating geophysical flows. However, robust and efficient
solvers for the resulting saddle-point and elliptic systems arising from
these discretizations continue to be an on-going challenge. One possible
approach for addressing this issue is to employ a method known as hybridization,
where the discrete equations are transformed such that classic static condensation
and local post-processing methods can be employed. However,
it is challenging to implement
hybridization as performant parallel code within complex models,
whilst maintaining separation of concerns between applications scientists
and software experts.
In this paper, we introduce a domain-specific abstraction within the
Firedrake finite element library that permits the rapid execution
of these hybridization techniques within a code-generating framework. The
resulting framework composes naturally with Firedrake's solver environment,
allowing for the implementation of hybridization and static condensation as
runtime-configurable preconditioners via the Python interface to PETSc, petsc4py.
We provide examples derived from second order elliptic problems and geophysical
fluid dynamics. In addition, we demonstrate that hybridization shows great
promise for improving the performance of solvers for mixed finite element
discretizations of equations related to large-scale geophysical flows.
\end{abstract}

%\copyrightstatement{}

\introduction  %% \introduction[modified heading if necessary]
\label{sec:introduction}
The development of simulation software is an increasingly important aspect of modern
scientific computing, in the geosciences in particular. Such software requires a vast range of knowledge spanning several
disciplines, ranging from applications expertise to mathematical analysis to high-performance computing and low-level
code optimization. Software projects developing automatic code generation systems have become
quite popular in recent years, as such systems help create a separation of concerns which
focuses on a particular complexity independent from the rest. This allows for agile collaboration between computer scientists with hardware and software expertise, computational scientists with numerical algorithm expertise, and domain scientists such as meteorologists, oceanographers and climate scientists.
Examples of such projects in the domain of finite element methods
include FreeFEM++ \citep{freefem++}, Sundance \citep{sundance}, the FEniCS Project \citep{fenics},
Feel++ \citep{feel++}, and Firedrake \citep{RHM+15}.

The finite element method (FEM) is a mathematically robust framework for computing numerical
solutions of partial differential equations (PDEs) that has become increasingly popular in fluids and solids models across the geosciences, with a formulation that is highly
amenable to code-generation techniques. A description of the weak formulation of the PDEs,
together with appropriate discrete function spaces, is enough to characterize the finite
element problem. Both the FEniCS and Firedrake projects employ the \emph{Unified Form Language}
(UFL) \citep{UFL} to specify the finite element integral forms and discrete spaces necessary
to properly define the finite element problem. UFL is a highly expressive domain-specific language
(DSL) embedded in Python, which provides the necessary abstractions for code generation systems.

There are classes of finite element discretizations resulting in discrete systems that can be
solved more efficiently by directly manipulating local tensors. For example, the static condensation
technique for the reduction of global finite element systems \citep{guyan, irons} produces smaller
globally-coupled linear systems by eliminating interior unknowns to arrive at an equation for
the degrees of freedom defined on cell-interfaces only. This procedure is analogous to the point-wise
elimination of variables used in staggered finite difference codes, such as the ENDGame dynamical core
\citep{melvin2010inherently,wood2014inherently} of the UK Meteorological Office (Met Office), but requires
the local inversion of finite element systems. For finite element discretizations of coupled equations
relevant to geophysical flows, the hybridization technique \citep{arnoldbrezzi,brezzifortin,cockburn}
introduces Lagrange multipliers enforcing certain continuity constraints. Local static condensation can
then be applied to the augmented system to produce a reduced equation for the multipliers. Methods of
this type are often accompanied by local post-processing techniques that produce superconvergent
approximations, or enhanced conservation properties \citep{bramble,cockburn2010projection,cockburn2009superconvergent}.
These procedures require invasive manual intervention during the equation assembly process in
intricate numerical code.

In this paper, we provide a simple yet effective high-level abstraction for localized dense linear algebra
on systems derived from finite element problems. Using embedded DSL technology, we provide a means to
enable the rapid development of hybridization and static condensation techniques within an automatic
code-generation framework. In other words, the main contribution of this paper is in solving the problem
of automatically translating from the mathematics of static condensation and hybridization to compiled code. This automated translation facilitates the separation of concerns between applications scientists and computational/computer scientists, and facilitates the automated optimization of compiled code.
This framework provides an environment for the development and testing of numerics relevant
to the Gung-Ho Project, an initiative by the UK Met Office in designing the next-generation atmospheric dynamical core
using mixed finite element methods \citep{melvin2018mixed}. Our work is implemented in the
Firedrake finite element library and the PETSc \citep{petsc1997,PETsc2016} solver library, accessed
via the Python interface petsc4py \citep{petsc4py}.

The rest of the paper is organized as follows. We introduce common notation used throughout the paper
in Section \ref{subsec:notation}. The embedded DSL, called ``Slate'', is introduced in Section \ref{sec:slate-DSL},
which allows concise expression of localized linear algebra operations on finite element tensors. We
provide some contextual examples for static condensation and hybridization in Section \ref{subsec:examples},
including a discussion on post-processing. We then outline in Section \ref{sec:PCs} how, by interpreting static
condensation techniques as a preconditioner, we can go further, and
automate many of the symbolic manipulations necessary for
hybridization and static condensation. We first demonstrate our
implementation on a
manufactured problem derived from a second-order elliptic equation, starting in Section \ref{sec:applications}.
The first example compares a hybridizable discontinuous Galerkin (HDG) method with an optimized continuous
Galerkin method.
Section \ref{subsec:williamson5} illustrates the composability and relative performance of hybridization for
compatible mixed methods applied to a semi-implicit discretization of the nonlinear rotating shallow water
equations. Our final example in Section \ref{subsec:GW} demonstrates time-step robustness of a
hybridizable solver for a compatible finite element discretization of a rotating linear Boussinesq
model. Conclusions follow in Section \ref{sec:conclusions}.

\subsection{Notation}\label{subsec:notation}
We begin by establishing notation used throughout this paper. Let $\mathcal{T}_h$ denote a tessellation of
$\Omega \subset \mathbb{R}^n$, the computational domain, consisting of polygonal elements $K$ associated
with a mesh size parameter $h$, and $\partial\mathcal{T}_h = \lbrace e \in \partial K : K \in \mathcal{T}_h\rbrace$
the set of facets of $\mathcal{T}_h$. The set of facets \emph{interior} to the domain $\Omega$ is denoted by
$\mathcal{E}_h^\circ \equiv \partial\mathcal{T}_h\setminus\partial\Omega$. Similarly, we denote the set of
\emph{exterior} facets as $\mathcal{E}_h^\partial = \partial\mathcal{T}_h \cap \partial\Omega$.
For brevity, we denote the finite element integral forms over $\mathcal{T}_h$ and any facet set
$\Gamma \subset \partial\mathcal{T}_h$ by
\begin{align}
	\cellinner{u, v}_{K} &= \int_{K} u \cdot v\,\mathrm{d}x, &
	\facetinner{u, v}_{e} &=
	\int_{e} u\cdot v\,\mathrm{d}s,\\
	\cellinner{u, v}_{\mathcal{T}_h} &= \sum_{K \in \mathcal{T}_h} \cellinner{u, v}_{K}, &
	\facetinner{u, v}_{\Gamma} &=
	\sum_{e \in \Gamma} \facetinner{u, v}_{e},
\end{align}
\noindent where $\cdot$ should be interpreted as standard multiplication for scalar functions or a dot product
for vector functions.

For any double-valued vector field $\boldsymbol{w}$ on a facet $e \in \partial\mathcal{T}_h$, we define the
jump of its normal component across $e$ by
\begin{equation}
		\jump{\boldsymbol{w}}_e =
		\begin{cases}
		\boldsymbol{w}|_{e^+}\cdot\boldsymbol{n}_{e^+} +
		\boldsymbol{w}|_{e^-}\cdot\boldsymbol{n}_{e^-}, &
		e \in \mathcal{E}_h^\circ \\
		\boldsymbol{w}|_e \cdot \boldsymbol{n}_e, &
		e \in \mathcal{E}_h^\partial
	\end{cases}
\end{equation}
where $+$ and $-$ denotes arbitrarily but globally defined sides of the facet.
Here, $\boldsymbol{n}_{e^+}$ and $\boldsymbol{n}_{e^-}$ are the unit normal vectors with respect
to the positive and negative sides of the facet $e$. Whenever the facet domain is clear by the context,
we omit the subscripts for brevity and simply write $\jump{\cdot}$.

\section{A system for localized algebra on finite element tensors}\label{sec:slate-DSL}
We present an expressive language for dense linear algebra on the elemental matrix systems arising from finite
element problems. The language, which we call \emph{Slate}, provides typical mathematical operations performed
on matrices and vectors, hence the input syntax is comparable to high-level linear algebra software such as
MATLAB. The Slate language provides basic abstract building blocks which can be used by a specialized compiler
for linear algebra to generate low-level code implementations.

Slate is heavily influenced by the Unified Form Language (UFL) \citep{UFL, fenics}, a DSL embedded in Python
which provides symbolic representations of finite element forms.  The expressions can be compiled by a
\emph{form compiler}, which translates UFL into low level code for the local assembly of a form over the cells
and facets of a mesh. In a similar manner, Slate expressions are compiled to low level code that performs the
requested linear algebra element-wise on a mesh.

\subsection{An overview of Slate}\label{subsec:slate-overview}
To clarify conventions and the scope of Slate, we start by considering a general form. Suppose we have a
finite element form:
\begin{equation}\label{eq:form}
	a(\boldsymbol{c}; \boldsymbol{v}) =
	\sum_{K \in \mathcal{T}_h} \int_{K} \mathcal{I}^{c}(\boldsymbol{c};
	\boldsymbol{v})\mathrm{d}x +
	\sum_{e \in \mathcal{E}_h^\circ}\int_{e} \mathcal{I}^{f,\circ}(\boldsymbol{c};
	\boldsymbol{v})\mathrm{d}s +
	\sum_{e \in \mathcal{E}_h^\partial}\int_{e} \mathcal{I}^{f,\partial}(\boldsymbol{c};
	\boldsymbol{v})\mathrm{d}s,
\end{equation}
where $\mathrm{d}x$ and $\mathrm{d}s$ denote appropriate integration measures. The integral
form in \eqref{eq:form} is uniquely determined by its lists (possibly of 0-length) of arbitrary
coefficient functions $\boldsymbol{c} = (c_0, \cdots, c_p)$ in the associated finite element spaces,
arguments $\boldsymbol{v} = (v_0, \cdots, v_q)$ describing any test or trial functions, and its integrand
expressions for each integral type: $\mathcal{I}^{c}$, $\mathcal{I}^{f,\circ}$, $\mathcal{I}^{f, \partial}$.
The form $a(\boldsymbol{c}; \boldsymbol{v})$ describes a finite element form \emph{globally} over the entire
problem domain. Here, we will consider the case where the integrand $\mathcal{I}^{f,\circ}(\boldsymbol{c}; \boldsymbol{v})$ can be decomposed into two independent parts for each facet $e$: one for the positive
restriction ($+$) and the negative restriction ($-$). 

The contribution of \eqref{eq:form} in each cell $K$ of the mesh $\mathcal{T}_h$ is simply
\begin{equation}\label{eq:cell-local-form}
	a(\boldsymbol{c}; \boldsymbol{v})|_{K} =
	\int_{K} \mathcal{I}^{c}(\boldsymbol{c}; \boldsymbol{v})|_K \mathrm{d}x
	+ \sum_{e \in \partial K \setminus \partial\Omega} \int_{e}
	\mathcal{I}^{f,\circ}(\boldsymbol{c}; \boldsymbol{v})|_e \mathrm{d}s
	+ \sum_{e \in \partial K \cap \partial\Omega} \int_{e}
	\mathcal{I}^{f,\partial}(\boldsymbol{c}; \boldsymbol{v})|_e \mathrm{d}s.
\end{equation}
We call \eqref{eq:cell-local-form} the \emph{cell-local} contribution of
$a(\boldsymbol{c}; \boldsymbol{v})$, with
\begin{equation}
a(\boldsymbol{c}; \boldsymbol{v}) =
\sum_{K \in \mathcal{T}_h} a(\boldsymbol{c}; \boldsymbol{v})|_{K}.
\end{equation}
Equation \eqref{eq:cell-local-form} produces an element tensor which is mapped into a global data structure. However, before doing so, one may want to produce a new local tensor by algebraically manipulating different element tensors.
This is precisely the job of Slate.

The Slate language consists of two primary abstractions for linear algebra:
\begin{enumerate}
	\item terminal element tensors corresponding to multi-linear integral forms (matrices, vectors, and scalars),
	or assembled data (coefficient vectors); and
	\item expressions consisting of operations on terminal tensors.
\end{enumerate}
The composition of binary and unary operations on terminal tensors produces a \emph{Slate expression}.
Such expressions can be composed with other Slate objects in arbitrary ways, resulting in concise representations
of complex operations on locally assembled arrays.

\paragraph{Terminal tensors: } In Slate, one associates a tensor with data on an element either by using a form,
or assembled coefficient data:
\begin{itemize}
	\item \verb|Tensor|($a(\boldsymbol{c}; \boldsymbol{v})$)\\
	associates a form, expressed in UFL, with its local element tensor:
	\begin{equation}
		A^{K} \leftarrow a(\boldsymbol{c}; \boldsymbol{v})|_{K},
		\text{ for all } K \in \mathcal{T}_h.
	\end{equation}
	The number of arguments $\boldsymbol{v}$ determine the \emph{rank} of
	\verb|Tensor|, i.e.~scalars, vectors, and matrices are produced from 0-forms,
	1-forms, and 2-forms\footnote{As with UFL, Slate is capable of abstractly
		representing arbitrary rank tensors. However, only rank $\leq 2$ tensors are
		typically used in most finite element applications and therefore we currently
		only generate code for those ranks.}
	respectively.
	\item \verb|AssembledVector|($f$)\\
	where $f$ is some finite element function. The result associates a function
	with its local coefficient vectors.
\end{itemize}

\paragraph{Symbolic linear algebra: } Slate supports typical binary and unary operations in linear algebra,
with a high-level syntax close to mathematics. At the time of this paper, these include:
\begin{itemize}
	\item \verb|A + B|, the addition of two equal shaped tensors.
	\item \verb|A * B|, a contraction over the last index of \verb|A| and the first index of \verb|B|.
	This is the usual multiplicative operation on matrices, vectors, and scalars.
	\item \verb|-A|, the additive inverse (negation) of a tensor.
	\item \verb|A.T|, the transpose of a tensor.
	\item \verb|A.inv|, the inverse of a square tensor.
	\item \verb|A.solve(B, decomposition="...")|, the result of solving a local linear system $AX = B$ for $X$,
	optionally specifying a direct factorization strategy.
	\item \verb|A.blocks[indices]|, where A is a tensor from a mixed finite element space, allows extraction
	of subblocks of the tensor indexed by field (slices are allowed). For example, if a matrix $A$ corresponds to
	the bilinear form $a: V \times W \rightarrow \mathbb{R}$, where $V = V_0 \times \cdots\times V_n$ and
	$W = W_0 \times \cdots\times W_m$ are product spaces consisting of finite element spaces
	$\{V_i\}_{i=0}^n$, $\{W_i\}_{i=0}^m$, then the cell-local tensors have the form:
	\begin{equation}\label{eq:blockmatrix}
		A^{K} =
		\begin{bmatrix}
		A^{K}_{0 0} & A^{K}_{0 1} & \cdots & A^{K}_{0 m} \\
		A^{K}_{1 0} & A^{K}_{1 1} & \cdots & A^{K}_{1 m} \\
		\vdots & \vdots & \ddots & \vdots \\
		A^{K}_{n 0} & A^{K}_{n 1} & \cdots & A^{K}_{n m}
		\end{bmatrix}.
	\end{equation}
	\noindent The associated submatrix of \eqref{eq:blockmatrix} with indices
	$\boldsymbol{i} = (\boldsymbol{p}, \boldsymbol{q})$,
	$\boldsymbol{p} = \lbrace p_1, \cdots, p_r\rbrace$,
	$\boldsymbol{q} = \lbrace q_1, \cdots, q_c\rbrace$, is
	\begin{equation}
		A^{K}_{\boldsymbol{p} \boldsymbol{q}} =
		\begin{bmatrix}
		A^{K}_{p_1 q_1} & \cdots & A^{K}_{p_1 q_c} \\
		\vdots & \ddots & \vdots \\
		A^{K}_{p_r q_1} & \cdots & A^{K}_{p_r q_c}
		\end{bmatrix}
		= A^{K}\!\verb|.blocks[|\boldsymbol{p}, \boldsymbol{q}\verb|]|,
	\end{equation}
	\noindent where $\boldsymbol{p} \subset \lbrace 1, \cdots, n\rbrace$,
	$\boldsymbol{q} \subset \lbrace 1, \cdots, m\rbrace$.
\end{itemize}
These building blocks may be arbitrarily composed, giving a symbolic
expression for the linear algebra we wish to perform on each cell
during assembly. They provide the necessary algebraic framework for a large
class of problems, some of which we present in this paper.

\begin{figure}[htbp]
	\centering
	\includegraphics[width=0.4\textwidth]{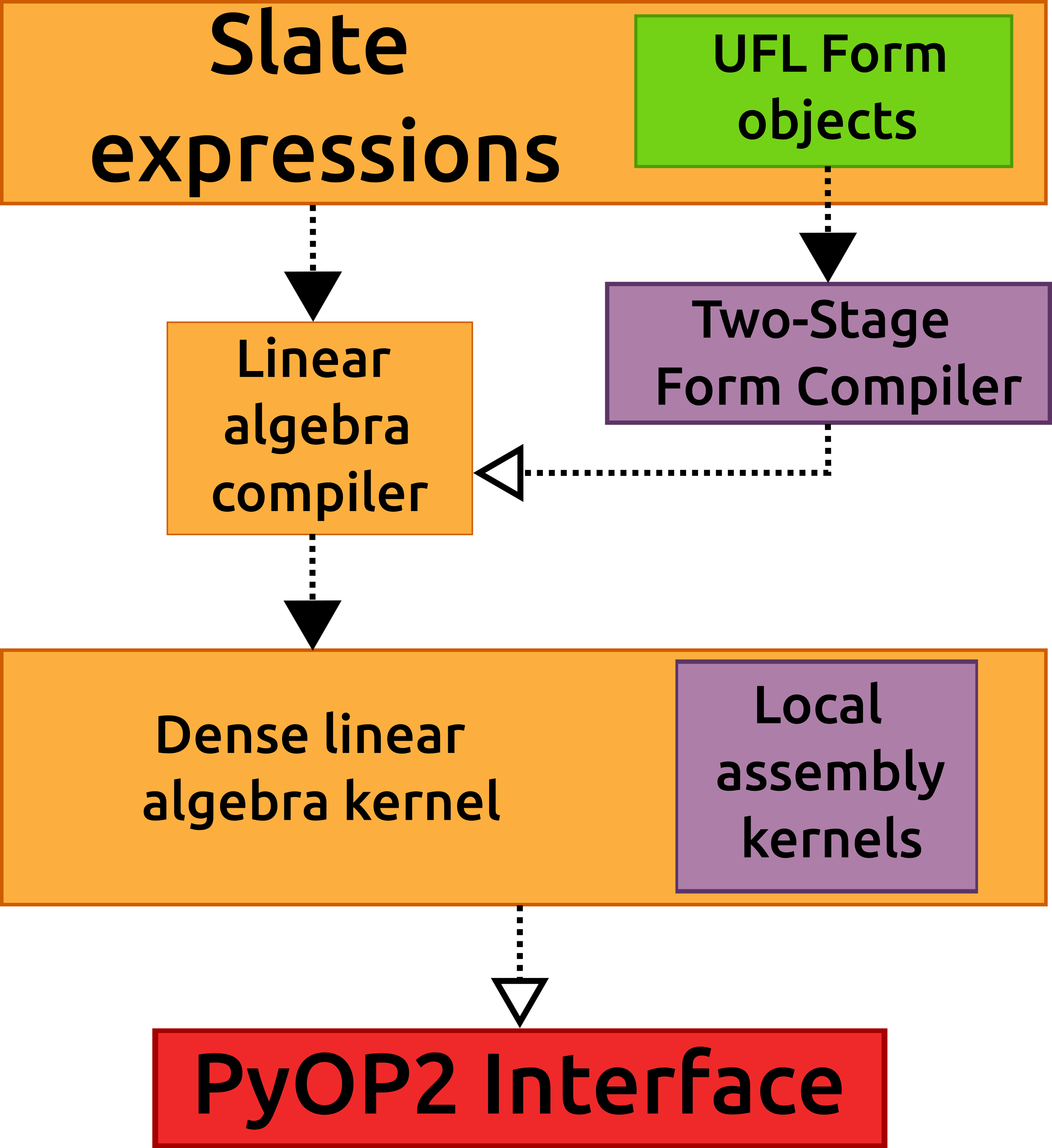}
	\caption{The Slate language wraps UFL objects describing the finite element
		system. The resulting Slate expressions are passed to a specialized linear
		algebra compiler, which produces a single ``macro" kernel assembling the local
		contributions and executes the dense linear algebra represented in Slate. The
		kernels are passed to the Firedrake's PyOP2 interface, which wraps the Slate kernel in
		a mesh-iteration kernel. Parallel scheduling, code generation, and compilation occurs
		after the PyOP2 layer.}
	\label{fig:slate-layer}
\end{figure}

In Firedrake, these \emph{Slate expressions} are transformed into
low-level code by a \emph{linear algebra compiler}. This uses the form compiler,
TSFC \citep{homolya2018tsfc}, to compile kernels for the assembly of terminal tensors and generates a dense
linear algebra kernel to be iterated cell-wise. At the time of this work, our compiler generates C++ code,
using the templated library Eigen \citep{eigen} for dense linear algebra. During execution, the local computations
in each cell are mapped into global data objects via appropriate indirection mappings using the PyOP2 framework
\citep{pyop2}. Figure \ref{fig:slate-layer} provides an illustration of the complete tool-chain.

\section{Examples}\label{subsec:examples}
We now present a few examples and discuss solution methods which require element-wise manipulations of finite
element systems and their specification in Slate. We stress here that Slate is not limited to these model problems;
rather these examples were chosen for clarity and to demonstrate key features of the Slate language. In Sections
\ref{sec:PCs} and \ref{sec:applications}, we discuss more intricate ways Slate is used in custom preconditioners.

For the hybridization of mixed and discontinuous Galerkin methods, we use a model elliptic equation.
Consider the second-order PDE with both Dirichlet and Neumann boundary conditions:
\begin{align}
	-\nabla\cdot\left(\kappa\nabla p\right) + c p
	&= f \text{ in } \Omega,\label{eq:helmholtzH1-a}\\
	p &= p_0 \text{ on } \partial\Omega_D,\label{eq:helmholtzH1-b}\\
	-\kappa\nabla p\cdot\boldsymbol{n} &= g \text{ on } \partial\Omega_N,
	\label{eq:helmholtzH1-c}
\end{align}
where $\partial\Omega_D \cup \partial\Omega_N = \partial\Omega$
and $\kappa$, $c: \Omega \rightarrow \mathbb{R}^{+}$ are positive-valued coefficients. Rewriting as a
first-order system, we obtain the mixed problem:
\begin{align}
	\mu\boldsymbol{u} + \nabla p &= 0 \quad \text{in }
	\Omega,
	\label{eq:mixedpde-ex-1}\\
	\nabla \cdot \boldsymbol{u} + c p &= f \quad
	\text{in } \Omega,
	\label{eq:mixedpde-ex-2}\\
	p &= p_0 \quad \text{on } \partial\Omega_D,
	\label{eq:mixedpde-ex-3}\\
	\boldsymbol{u}\cdot\boldsymbol{n} &= g \quad \text{on } \partial\Omega_N,
	\label{eq:mixedpde-ex-4}
\end{align}
where $\mu = \kappa^{-1}$ and $\boldsymbol{u} = -\kappa\nabla p$ is the velocity variable.

\subsection{Hybridization of mixed methods}\label{subsec:hybrid-mixed-example}
To motivate our discussion in this section, we start by recalling the mixed method for
\eqref{eq:mixedpde-ex-1}--\eqref{eq:mixedpde-ex-4}. Methods of this type seek approximations
$(\boldsymbol{u}_h, p_h)$ in finite-dimensional subspaces
$\boldsymbol{U}_h \times V_h \subset \boldsymbol{H}(\text{div})\times L^2$, defined by:
\begin{align}\label{eq:mixed-spaces}
	\boldsymbol{U}_h &=
	\lbrace \boldsymbol{w} \in \boldsymbol{H}(\text{div}; \Omega)
	: \boldsymbol{w}|_K \in \boldsymbol{U}(K), \forall K \in \mathcal{T}_h,
	\boldsymbol{w}\cdot\boldsymbol{n} = g \text{ on } \partial\Omega_N \rbrace,\\
	V_h &=
	\lbrace \phi \in L^2(\Omega) : \phi|_K \in V(K), \forall K \in \mathcal{T}_h\rbrace.
\end{align}
The space $\boldsymbol{U}_h$ consists of $\boldsymbol{H}(\text{div})$-conforming piecewise vector polynomials,
where choices of $\boldsymbol{U}(K)$ typically include the Raviart-Thomas (RT),
Brezzi-Douglas-Marini (BDM), or Brezzi-Douglas-Fortin-Marini (BDFM) elements
\citep{mixiedFEMfor2ndOrdPrb, twofamiliesofmixedFEM, nedelec, raviartthomas}.
$V_h$ is the Lagrange family of discontinuous polynomials. These spaces are of particular interest when simulating
geophysical flows, since choosing the right pairing results in stable discretizations with desirable conservation
properties and avoids spurious computational modes. We refer the reader to
\citet{cottershipton,cotterthuburn,natale2016compatible,shipton2018higher} for a discussion of mixed
methods relevant for geophysical fluid dynamics. Two examples of such a discretization is presented in
Section \ref{subsec:williamson5}.

The mixed finite element formulation of \eqref{eq:mixedpde-ex-1}--\eqref{eq:mixedpde-ex-4}
reads as follows: find $(\boldsymbol{u}_h, p_h) \in \boldsymbol{U}_{h} \times V_h$ satisfying
\begin{align}
	\cellinner{\boldsymbol{w}, \mu\boldsymbol{u}_h}_{\mathcal{T}_h} -
	\cellinner{\nabla\cdot\boldsymbol{w}, p_h}_{\mathcal{T}_h}
	&= -\facetinner{\boldsymbol{w}\cdot\boldsymbol{n}, p_0}_{\partial\Omega_D},
	&\forall \boldsymbol{w} \in \boldsymbol{U}_{h, 0},
	\label{eq:mixed-mth-A}\\
	\cellinner{\phi, \nabla\cdot\boldsymbol{u}_h}_{\mathcal{T}_h} +
	\cellinner{\phi, c p_h}_{\mathcal{T}_h} &= \cellinner{\phi, f}_{\mathcal{T}_h},
	&\forall \phi \in V_h,
	\label{eq:mixed-mth-B}
\end{align}
where $\boldsymbol{U}_{h, 0}$ is the space of functions in $\boldsymbol{U}_h$
whose normal components vanish on $\partial\Omega_N$. The discrete system
is obtained by first expanding the solutions in terms of the finite element
bases:
\begin{equation}
	\boldsymbol{u}_h = \sum_{i=1}^{N_{\boldsymbol{u}}} U_i \boldsymbol{\Psi}_i, \quad
	p_h = \sum_{i=1}^{N_p} P_i \xi_i,
\end{equation}
where $\lbrace \boldsymbol{\Psi}_i \rbrace_i$ and $\lbrace \xi_i \rbrace_i$ are
bases for $\boldsymbol{U}_h$ and $V_h$ respectively. Here, $U_i$ and $P_i$ are
the coefficients to be determined. As per standard Galerkin-based finite element methods,
taking $\boldsymbol{w} = \boldsymbol{\Psi}_j$, $j \in \lbrace 1, \cdots N_{\boldsymbol{u}} \rbrace$
and $\phi = \xi_j$, $j \in \lbrace 1, \cdots N_p \rbrace$ in
\eqref{eq:mixed-mth-A}--\eqref{eq:mixed-mth-B}
produces the discrete saddle point system:
\begin{equation}\label{eq:indefinitesys}
	\begin{bmatrix}
		A & -B^T \\
		B & C
	\end{bmatrix}
	\begin{Bmatrix}
		U \\
		P
	\end{Bmatrix} =
	\begin{Bmatrix}
		F_{0} \\
		F_{1}
	\end{Bmatrix}.
\end{equation}
where $U = \lbrace U_i \rbrace$, $P = \lbrace P_i \rbrace$ are the coefficient vectors, and
\begin{align}
A_{ij} &= \cellinner{\boldsymbol{\Psi}_i, \mu\boldsymbol{\Psi}_j}_{\mathcal{T}_h}, \\
B_{ij} &= \cellinner{\xi_i, \nabla\cdot\boldsymbol{\Psi}_j}_{\mathcal{T}_h},\\
C_{ij} &= \cellinner{\xi_i, c \xi_j}_{\mathcal{T}_h},\\
F_{0,j} &= -\facetinner{\boldsymbol{\Psi}_j\cdot\boldsymbol{n}, p_0}_{\partial\Omega_D},\\
F_{1,j} &= \cellinner{\xi_j, f}_{\mathcal{T}_h}.
\end{align}

Methods to efficiently invert such systems include $\boldsymbol{H}(\text{div})$-multigrid
\citep{arnoldfalkwinther2000} (requiring complex overlapping-Schwarz smoothers), global
Schur-complement factorizations (which require an approximation to the inverse of the
\emph{dense}\footnote{The Schur-complement, while elliptic, is globally dense
	due to the fact that $A$ has a dense inverse. This is a result of velocities in $\boldsymbol{U}_h$
	having continuous normal components across cell-interfaces.}
elliptic Schur-complement $C + B A^{-1} B^T$), or auxiliary space multigrid \citep{hiptmairxu2007}.
Here, we focus on a solution approach using a hybridized mixed method \citep{arnoldbrezzi, brezzifortin}.

The hybridization technique replaces the original system with a discontinuous variant, decoupling
the velocity degrees of freedom between cells. This is done by replacing the discrete solution space
for $\boldsymbol{u}_h$ with the ``broken'' space $\boldsymbol{U}^d_h$, defined as:
\begin{equation}\label{eq:nc-hdiv-space}
	\boldsymbol{U}^d_h =
	\lbrace \boldsymbol{w} \in \lbrack L^2(\Omega)\rbrack^n :
	\boldsymbol{w}|_K \in \boldsymbol{U}(K), \forall K \in \mathcal{T}_h\rbrace.
\end{equation}
The vector finite element space $\boldsymbol{U}^d_h$ is a subspace of $\lbrack L^2(\Omega) \rbrack^n$
consisting of local $\boldsymbol{H}(\text{div})$ functions, but normal components are no longer
continuous on $\partial\mathcal{T}_h$. The approximation space for $p_h$ remains unchanged.

Next, Lagrange multipliers are introduced as an auxiliary variable in the space $M_h$,
defined only on cell-interfaces:
\begin{equation}\label{eq:trace-space}
	M_h = \lbrace \gamma \in L^2(\partial\mathcal{T}_h) :
	\gamma|_e \in M(e), \forall e \in \partial\mathcal{T}_h\rbrace,
\end{equation}
\noindent where $M(e)$ denotes a polynomial space defined on each facet. We
call $M_h$ the space of approximate traces. Functions in $M_h$ are discontinuous
across vertices in two-dimensions, and vertices/edges in three-dimensions.

Deriving the hybridizable mixed system is accomplished through integration by parts over each element $K$.
Testing with $\boldsymbol{w} \in \boldsymbol{U}_h^d(K)$ and integrating \eqref{eq:mixedpde-ex-1} over
$K$ produces:
\begin{equation}
	\cellinner{\boldsymbol{w}, \mu \boldsymbol{u}^d_h}_K -
	\cellinner{\nabla\cdot\boldsymbol{w}, p_h}_K +
	\facetinner{\boldsymbol{w}\cdot\boldsymbol{n}, \lambda_h}_{\partial K}
	= -\facetinner{\boldsymbol{w}\cdot\boldsymbol{n}p_0}_{\partial K \cap \partial\Omega_D}.
\end{equation}
The trace function $\lambda_h$ is introduced in surface integrals and approximates $p_h$ on elemental
boundaries. An additional constraint equation is added to close the system. The resulting hybridizable
formulation reads: find $(\boldsymbol{u}^d_h, p_h, \lambda_h) \in \boldsymbol{U}^d_h\times V_h \times M_{h}$
such that
\begin{align}
	\cellinner{\boldsymbol{w}, \mu \boldsymbol{u}_h^d}_{\mathcal{T}_h}
	- \cellinner{\nabla\cdot\boldsymbol{w}, p_h}_{\mathcal{T}_h}
	+ \facetinner{\jump{\boldsymbol{w}}, \lambda_h}_{\partial\mathcal{T}_h \setminus \partial\Omega_D}
	& = -\facetinner{\boldsymbol{w}\cdot\boldsymbol{n}, p_0}_{\partial\Omega_D},
	&\forall \boldsymbol{w} \in \boldsymbol{U}_h^d, \label{eq:hybridmixed-eq1}\\
	\cellinner{\phi, \nabla\cdot\boldsymbol{u}_h^d}_{\mathcal{T}_h}
	+ \cellinner{\phi, c p_h}_{\mathcal{T}_h} &=
	\cellinner{\phi, f}_{\mathcal{T}_h},
	&\forall \phi \in V_h, \label{eq:hybridmixed-eq2}\\
	\facetinner{\gamma, \jump{\boldsymbol{u}_h^d}}_{\partial\mathcal{T}_h\setminus\partial\Omega_D}
	&= \facetinner{\gamma, g}_{\partial\Omega_N},
	&\forall \gamma \in M_{h, 0},\label{eq:hybridmixed-eq3}
\end{align}
where $M_{h, 0}$ denotes the space of traces vanishing on $\partial\Omega_D$.
The constraint in \eqref{eq:hybridmixed-eq3} enforces both continuity of the normal components
of $\boldsymbol{u}^d_h$ across elemental boundaries, as well as the boundary condition on $\partial\Omega_N$.
If the space of Lagrange multipliers $M_{h}$ is chosen appropriately, then the ``broken'' velocity
$\boldsymbol{u}^d_h$, albeit sought a priori in a discontinuous space, will coincide with its
$\boldsymbol{H}(\text{div})$-conforming counterpart. Specifically, the formulations in
\eqref{eq:hybridmixed-eq1}--\eqref{eq:hybridmixed-eq2} and
\eqref{eq:mixed-mth-A}--\eqref{eq:mixed-mth-B} are solving equivalent problems
if the normal components of $\boldsymbol{w} \in \boldsymbol{U}_h$ lie in the same polynomial space as the
trace functions \citep{arnoldbrezzi}.

The discrete matrix system arising from \eqref{eq:hybridmixed-eq1}--\eqref{eq:hybridmixed-eq3} has the
general form:
\begin{equation}\label{eq:hybrid-sys}
	\begin{bmatrix}
		A_{00} & A_{01} & A_{02} \\
		A_{10} & A_{11} & A_{12} \\
		A_{20} & A_{21} & A_{22}
	\end{bmatrix}
	\begin{Bmatrix}
		U^d\\
		P\\
		\Lambda
	\end{Bmatrix} =
	\begin{Bmatrix}
		F_0 \\
		F_1 \\
		F_2
	\end{Bmatrix},
\end{equation}
where the discrete system is produced by expanding functions in terms of the finite element
bases for $\boldsymbol{U}^d_h$, $V_h$, and $M_{h}$ like before.
Upon initial inspection, it may not appear to be advantageous to replace our original
formulation with this augmented equation-set; the hybridizable system has substantially
more total degrees of freedom. However, \eqref{eq:hybrid-sys} has a considerable
advantage over \eqref{eq:indefinitesys} in the following ways:
\begin{enumerate}
	\item Since both $\boldsymbol{U}_h^d$ and $V_h$ are discontinuous spaces, $U^d$ and $P$
	are coupled only within the cell. This allows us to simultaneously eliminate
	both unknowns via \emph{local} static condensation to produce a significantly smaller
	global (hybridized) problem for the trace unknowns, $\Lambda$:
	\begin{equation}
		S\Lambda = E,
		\label{eq:lambda-sys}
	\end{equation}
	where $S = \lbrace S^K\rbrace_{K \in \Omega_h}$ and $E = \lbrace E^K\rbrace_{K \in \Omega_h}$ are
	assembled by gathering the local contributions:
	\begin{align}
		S^K &= A^K_{22} -
		\begin{bmatrix}
			A^K_{20} & A^K_{21}
		\end{bmatrix}
		\begin{bmatrix}
			A^K_{0 0} & A^K_{0 1} \\
			A^K_{1 0} & A^K_{1 1}
		\end{bmatrix}^{-1}
		\begin{bmatrix}
			A^K_{02} \\
			A^K_{12}
		\end{bmatrix}
		\label{eq:hybrid-S},\\
		E^K &= F^K_2 -
		\begin{bmatrix}
		A^K_{20} & A^K_{21}
		\end{bmatrix}
		\begin{bmatrix}
		A^K_{0 0} & A^K_{0 1} \\
		A^K_{1 0} & A^K_{1 1}
		\end{bmatrix}^{-1}
		\begin{Bmatrix}
			F^K_0 \\
			F^K_1
		\end{Bmatrix}
		\label{eq:hybrid-E}.
	\end{align}
	Note that the inverse of the block matrix in \eqref{eq:hybrid-S} and \eqref{eq:hybrid-E}
	is \emph{never} evaluated globally; the elimination can be performed locally by performing
	a sequence of Schur-complement reductions within each cell.
	\item The matrix $S$ is sparse, symmetric, positive-definite, and spectrally similar
	to the dense Schur-complement $C + B A^{-1} B^T$ from \eqref{eq:indefinitesys} of the original
	mixed formulation \citep{cockburn}.
	\item Once $\Lambda$ is computed, both $U^d$ and $P$ can be recovered
	locally in each element. This can be accomplished in a number ways.
	One way is to compute $P^K$ by solving:
	\begin{equation}\label{eq:hybrid-p-sys}
		\left(A^K_{11} - A^K_{10} \left(A_{00}^K\right)^{-1} A^K_{01}\right)P^K = F^K_1 - A^K_{10}  \left(A_{00}^K\right)^{-1} F^K_0
		- \left(A^K_{12} - A^K_{10}  \left(A_{00}^K\right)^{-1} A^K_{02}\right)\Lambda^K,
	\end{equation}
	followed by solving for $\left(U^d\right)^K$:
	\begin{equation}\label{eq:hybrid-u-sys}
		A_{00}^K \left(U^d\right)^K = F^K_0 - A^K_{01} P^K - A^K_{02}\Lambda^K.
	\end{equation}
	Similarly, one could rearrange the order in which each variable is reconstructed.
	\item If desired, the solutions can be improved further through local
	post-processing. We highlight two such procedures, for $U^d$ and $P$ respectively,
	in Section \ref{subsec:postprocessing}.
\end{enumerate}

Listing \ref{lst:slateex-scsys} displays the corresponding Slate code for assembling the trace
system, solving \eqref{eq:lambda-sys}, and recovering the eliminated unknowns. For a complete
reference on how to formulate the hybridized mixed system 
\eqref{eq:hybridmixed-eq1}--\eqref{eq:hybridmixed-eq3} in UFL, we refer the reader to \citet{UFL}.
We remark that, in the case of this hybridizable system, \eqref{eq:hybrid-sys} contains zero-valued
blocks which can simplify the resulting expressions in \eqref{eq:hybrid-S}--\eqref{eq:hybrid-E}
and \eqref{eq:hybrid-p-sys}--\eqref{eq:hybrid-u-sys}. This is not true in general and therefore
the expanded form using all sub-blocks of \eqref{eq:hybrid-sys} is presented for completeness.

\begin{minipage}[!htp]{\linewidth}
\centering
\begin{lstlisting}[style=framed, frame=single, label={lst:slateex-scsys}, language={[firedrake]python},
caption={Firedrake code for solving \eqref{eq:hybrid-sys} via static condensation
and local recovery, given UFL expressions \texttt{a}, \texttt{L} for
\eqref{eq:hybridmixed-eq1}--\eqref{eq:hybridmixed-eq3}. Arguments of
the mixed space $\boldsymbol{U}^d_h\times V_h\times M_h$ are indexed by 0, 1, and 2
respectively. Lines \ref{eq:buildS} and \ref{eq:buildE} are symbolic expressions
for \eqref{eq:hybrid-S} and \eqref{eq:hybrid-E} respectively. Any vanishing conditions on the
trace variables should be provided as boundary conditions during operator assembly
(line \ref{eq:assembleS-code}).
Lines \ref{eq:recover-ph} and \ref{eq:recover-uh} are expressions
for \eqref{eq:hybrid-p-sys} and \eqref{eq:hybrid-u-sys} (using LU). Code generation
occurs in lines \ref{eq:assembleS-code}, \ref{eq:assembleE-code}, \ref{eq:assemble-ph},
and \ref{eq:assemble-uh}. A global linear solver for the reduced system is created and used
in line \ref{eq:solve-code}. Configuring the linear solver is done by providing an appropriate Python
dictionary of solver options for the PETSc library.
}]
# Element tensors defining the local 3-by-3 block system
_A = Tensor(a)
_F = Tensor(L)

# Extracting blocks for Slate expression of the reduced system
A = _A.blocks
F = _F.blocks
@\label{eq:buildS}@S = A[2, 2] - A[2, :2] * A[:2, :2].inv * A[:2, 2]
@\label{eq:buildE}@E = F[2] - A[2, :2] * A[:2, :2].inv * F[:2]

# Assemble and solve: @\color{DarkGreen}$S\Lambda = E$@
@\label{eq:assembleS-code}@Smat = assemble(S, bcs=[...])
@\label{eq:assembleE-code}@Evec = assemble(E)
lambda_h = Function(M)
@\label{eq:solve-code}@solve(Smat, lambda_h, Evec, solver_parameters={"ksp_type": "preonly",
                                               "pc_type": "lu"})
p_h = Function(V)                              # Function to store the result: @\color{DarkGreen}$P$@
u_h = Function(U)                              # Function to store the result: @\color{DarkGreen}$U^d$@

# Intermediate expressions
Sd = A[1, 1] - A[1, 0] * A[0, 0].inv * A[0, 1]
Sl = A[1, 2] - A[1, 0] * A[0, 0].inv * A[0, 2]
Lambda = AssembledVector(lambda_h)             # Local coefficient vector for @\color{DarkGreen}$\Lambda$@
P = AssembledVector(p_h)                       # Local coefficient vector for @\color{DarkGreen}$P$@

# Local solve expressions for @\color{DarkGreen}$P$@ and @\color{DarkGreen}$U^d$@
@\label{eq:recover-ph}@p_sys = Sd.@\color{BurntOrange}solve@(F[1] - A[1, 0] * A[0, 0].inv * F[0] - Sl * Lambda,
                 decomposition="PartialPivLu")
@\label{eq:recover-uh}@u_sys = A[0, 0].@\color{BurntOrange}solve@(F[0] - A[0, 1] * P - A[0, 2] * Lambda,
                      decomposition="PartialPivLu")
@\label{eq:assemble-ph}@assemble(p_sys, p_h)
@\label{eq:assemble-uh}@assemble(u_sys, u_h)
\end{lstlisting}
\end{minipage}

\subsection{Hybridization of discontinuous Galerkin methods}\label{subsubsec:HDG}
The hybridized discontinuous Galerkin (HDG) method is a natural extension of discontinuous
Galerkin (DG) discretizations. Here, we consider a specific HDG discretization, namely the
LDG-H method \citep{cockburn2010projection}. Other forms of HDG that involve local lifting
operators can also be implemented in this software framework by the introduction of additional
local (i.e., discontinuous) variables in the definition of the local solver.

To construct the LDG-H discretization, we define the DG numerical fluxes $\widehat{p}$
and $\widehat{\boldsymbol{u}}$ to be functions of the trial unknowns and a new
independent unknown in the trace space $M_h$:
\begin{align}
	\widehat{\boldsymbol{u}}(\boldsymbol{u}_h, p_h, \lambda_h; \tau) &=
	\boldsymbol{u}_h + \tau\left(p_h - \widehat{p}\right)\boldsymbol{n},
	\label{eq:num-fluxes-u} \\
	\widehat{p}(\lambda_h) &= \lambda_h
	\label{eq:num-fluxes-p},
\end{align}
where $\lambda_h \in M_{h}$ is a function approximating the trace of $p$ on $\partial\mathcal{T}_h$ and
$\tau$ is a positive function that may vary on each facet $e \in \partial\mathcal{T}_h$.
The full LDG-H formulation reads as follows. Find
$(\boldsymbol{u}_h, p_h, \lambda_h) \in \boldsymbol{U}_h \times V_h \times M_{h}$
such that
\begin{align}
	\cellinner{\boldsymbol{w}, \mu\boldsymbol{u}_h}_{\mathcal{T}_h} -
	\cellinner{\nabla\cdot\boldsymbol{w}, p_h}_{\mathcal{T}_h} +
	\facetinner{\jump{\boldsymbol{w}}, \lambda_h}_{\partial\mathcal{T}_h} &= 0,
	&\forall \boldsymbol{w} \in \boldsymbol{U}_h,
	\label{eq:hdg-fem-A}\\
	-\cellinner{\nabla \phi, \boldsymbol{u}_h}_{\mathcal{T}_h} +
	\facetinner{\phi, \jump{\boldsymbol{u}_h + \tau\left(p_h - \lambda_h\right)\boldsymbol{n}}
	}_{\partial\mathcal{T}_{h}}
	+ \cellinner{\phi, c p_h}_{\mathcal{T}_h} &= \cellinner{\phi, f}_{\mathcal{T}_h},
	&\forall \phi \in V_h,
	\label{eq:hdg-fem-B}\\
	\facetinner{\gamma,\jump{\boldsymbol{u}_h + \tau\left(p_h - \lambda_h\right)\boldsymbol{n}}
	}_{\partial\mathcal{T}_h\setminus\partial\Omega_D} + \facetinner{\gamma, \lambda_h}_{\partial\Omega_D}
	&= \facetinner{\gamma, g}_{\partial\Omega_N} + \facetinner{\gamma, p_0}_{\partial\Omega_D},
	&\forall \gamma \in M_h,
	\label{eq:hdg-fem-C}
\end{align}
Equation \eqref{eq:hdg-fem-C} enforces continuity of the numerical flux
$\widehat{\boldsymbol{u}}$ on $\partial\mathcal{T}_h$, which in turn produces a flux
that is single-valued on the facets. Note that the choice of $\tau$ has a significant
influence on the expected convergence rates of the computed solutions.

The matrix system arising from \eqref{eq:hdg-fem-A}--\eqref{eq:hdg-fem-C} has the same
general form as that of the hybridized mixed method in \eqref{eq:hybrid-sys}, except
all sub-blocks are now populated with non-zero entries due to the coupling of trace
functions with both $p_h$ and $\boldsymbol{u}_h$. However, all previous properties
of the discrete matrix system from Section \ref{subsec:hybrid-mixed-example} still
apply. The Slate expressions for the local elimination and reconstruction
operations will be identical to those of Listing \ref{lst:slateex-scsys}. For the interested
reader, a unified analysis of hybridization methods (both mixed and DG) for second-order
elliptic equations is presented in \cite{cockburn,cockburn2016static}.

\subsection{Local post-processing}\label{subsec:postprocessing}
For both mixed \citep{arnoldbrezzi,bramble,stenberg} and discontinuous
Galerkin methods
\citep{cockburn2010projection,cockburn2009superconvergent}, it is
possible to locally post-process solutions to obtain superconvergent
approximations (gaining one order of accuracy over the unprocessed
solution). These methods can be expressed as local solves on each
element, and so, in addition to static condensation,
the Slate language also provides access to code generation for local
post-processing of computed solutions.

Here, we present two post-processing techniques: one for scalar fields, and another for
the vector unknown. The Slate code follows naturally from previous discussions in
Sections \ref{subsec:hybrid-mixed-example} and \ref{subsubsec:HDG}, using the standard set
of operations on local tensors summarized in Section \ref{subsec:slate-overview}.

\subsubsection{Post-processing of the scalar solution}
Our first example is a modified version of the procedure presented by \citet{stenberg} for
enhancing the accuracy of the scalar solution. This was also highlighted within the context
of hybridizing eigenproblems by \citet{cockburn2010hybridization}. This post-processing
technique can be used for both the hybridized mixed and LDG-H methods.

Let $\mathcal{P}_{k}(K)$ denote a polynomial space of degree $\leq k$ on an
element $K \in \mathcal{T}_h$. Then for a given pair of computed solutions
$\boldsymbol{u}_h, p_h$ of the hybridized methods, we define
the post-processed scalar $p_h^\star \in \mathcal{P}_{k+1}(K)$ as the unique
solution of the local problem:
\begin{align}
	\cellinner{\nabla w, \nabla p_h^\star}_{K} &=
	-\cellinner{\nabla w, \kappa^{-1}\boldsymbol{u}_h}_{K},
	&\forall w \in \mathcal{P}_{k+1}^{\perp,l}(K),
	\label{eq:scalarpp-a}\\
	\cellinner{v, p_h^\star}_{K} &= \cellinner{v, p_h}_{K},
	&\forall v \in \mathcal{P}_{l}(K),
	\label{eq:scalarpp-b}
\end{align}
where $0 \leq l \leq k$. Here, the space $\mathcal{P}_{k+1}^{\perp,l}(K)$
denotes the $L^2$-orthogonal complement of $\mathcal{P}_l(K)$. This
post-processing method directly uses the definition of the flux
$\boldsymbol{u}_h = - \kappa\nabla p_h$ to
construct the local problem. In practice, the space $\mathcal{P}_{k+1}^{\perp,l}(K)$
may be constructed using an orthogonal hierarchical basis, and solving
\eqref{eq:scalarpp-a}--\eqref{eq:scalarpp-b} amounts to inverting a local symmetric
positive definite system.

At the time of this work, Firedrake does not
support the construction of such a finite element basis. However,
we can introduce Lagrange multipliers to enforce the orthogonality constraint.
The resulting local problem then becomes the following mixed system: find
$(p_h^\star, \psi) \in \mathcal{P}_{k+1}(K)\times \mathcal{P}_l(K)$
such that
\begin{align}
	\cellinner{\nabla w, \nabla p_h^\star}_{K} + \cellinner{w, \psi}_{K} &=
	-\cellinner{\nabla w, \kappa^{-1}\boldsymbol{u}_h}_{K},
	&\forall w \in \mathcal{P}_{k+1}(K),
	\label{eq:mixedscalarpp-a}\\
	\cellinner{\phi, p_h^\star}_{K} &= \cellinner{\phi, p_h}_{K},
	&\forall \phi \in \mathcal{P}_l(K),
	\label{eq:mixedscalarpp-b}
\end{align}
where $0 \leq l \leq k$. The local problems \eqref{eq:mixedscalarpp-a}--\eqref{eq:mixedscalarpp-b} and
\eqref{eq:scalarpp-a}--\eqref{eq:scalarpp-b} are equivalent, with the Lagrange
multiplier $\psi$ enforcing orthogonality of test functions in $\mathcal{P}_{k+1}(K)$
with functions in $\mathcal{P}_l(K)$.

This post-processing method produces a new approximation which
superconverges at a rate of $k+2$ for hybridized mixed methods
\citep{arnoldbrezzi,cockburn2010hybridization,stenberg}. For the LDG-H
method, $k+2$ superconvergence is achieved when
$\tau = \mathcal{O}(1)$ and $\tau = \mathcal{O}(h)$, but only $k+1$
convergence is achieved when $\tau = \mathcal{O}(1/h)$
\citep{cockburn2010projection,cockburn2009superconvergent}. We demonstrate the increased
accuracy in computed solutions in Section \ref{subsec:HDGCGcomp}. An
abridged example using Firedrake and Slate is provided in Listing \ref{lst:slateex-spp}.

\begin{minipage}[!htp]{\linewidth}
\centering
\begin{lstlisting}[style=framed, frame=single, label={lst:slateex-spp}, language={[firedrake]python},
caption={Example of local post-processing using Firedrake and Slate. Here, we locally solve
the mixed system defined in \eqref{eq:scalarpp-a}--\eqref{eq:scalarpp-b}. The corresponding
symbolic local tensors are defined in lines \ref{eq:scalar-pp-lhs} and \ref{eq:scalar-pp-rhs}. The
Slate expression for directly inverting the local system is written in line \ref{eq:ppsolve-expr}.
In line \ref{eq:scalar-pp-slate}, a Slate-generated kernel is produced which solves the resulting
linear system in each cell.
Since we are not interested in the multiplier, we only
return the block corresponding to the new pressure field.}]
# Define spaces for the higher-order pressure approximation and Lagrange multipliers
DGk1 = FunctionSpace(mesh, "DG", degree + 1)
DG0 = FunctionSpace(mesh, "DG", 0)
W = DGk1 * DG0
p, psi = TrialFunctions(W)
w, phi = TestFunctions(W)

# Create local Slate tensors for the post-processing system
@\label{eq:scalar-pp-lhs}@K = Tensor((inner(grad(p), grad(w)) + inner(psi, w) + inner(p, phi))*dx)
# Use computed pressure @\color{DarkGreen}$p_h$@ and flux @\color{DarkGreen}$\boldsymbol{u}_h$@ in right-hand side
@\label{eq:scalar-pp-rhs}@F = Tensor((-inner(u_h, grad(w)) + inner(p_h, phi))*dx)
@\label{eq:ppsolve-expr}@E = K.inv * F

# Function for the post-processed scalar field @\color{DarkGreen}$p_h^\star$@
p_star = Function(DGk1, name="Post-processed scalar")
@\label{eq:scalar-pp-slate}@assemble(E.blocks[0], p_star)      # Only want the first field (pressure)
\end{lstlisting}
\end{minipage}

\subsubsection{Post-processing of the flux}
Our second example illustrates a procedure that uses the numerical flux of an HDG
discretization for \eqref{eq:mixedpde-ex-1}--\eqref{eq:mixedpde-ex-4}. Within
the context of the LDG-H method, we can use the numerical trace in \eqref{eq:num-fluxes-u}
to produce a vector field that is $\boldsymbol{H}(\text{div})$-conforming. The technique
we outline here follows that of \citet{cockburn2009superconvergent}.

Let $\mathcal{T}_h$ be a mesh consisting of simplices\footnote{
	This particular post-processing strategy only works on triangles and tetrahedra.}.
On each element
$K \in \mathcal{T}_h$, we define a new function $\boldsymbol{u}_h^\star$ to
be the unique element of the local Raviart-Thomas space
$\lbrack \mathcal{P}_k(K)\rbrack^n + \boldsymbol{x}\mathcal{P}_k(K)$ satisfying
\begin{align}
	\cellinner{\boldsymbol{r}, \boldsymbol{u}_h^\star}_{K} &=
	\cellinner{\boldsymbol{r}, \boldsymbol{u}_h}_{K},
	&\forall \boldsymbol{r} \in \lbrack \mathcal{P}_{k-1}(K)\rbrack^n,
	\label{eq:fluxpp-a}\\
	\facetinner{\mu, \boldsymbol{u}_h^\star\cdot\boldsymbol{n}}_{e} &=
	\facetinner{\mu, \widehat{\boldsymbol{u}}_h\cdot\boldsymbol{n}}_{e},
	&\forall \mu \in \mathcal{P}_k(e), \forall e \in \partial K.
	\label{eq:fluxpp-b}
\end{align}
This local problem produces a new velocity $\boldsymbol{u}_h^\star$ with the following properties:
\begin{enumerate}
	\item $\boldsymbol{u}_h^\star$ converges at the \emph{same} rate as $\boldsymbol{u}_h$
	for all choices of $\tau$ producing a solvable system for
	\eqref{eq:hdg-fem-A}--\eqref{eq:hdg-fem-C}. However,
	\item $\boldsymbol{u}_h^\star \in \boldsymbol{H}(\text{div}; \Omega)$. That is,
	\begin{equation}
		\jump{\boldsymbol{u}_h^\star}_e = 0, \quad\forall e \in
		\mathcal{E}_h^\circ.
	\end{equation}
	\item Additionally, the divergence of $\boldsymbol{u}_h^\star$ convergences at a rate of
	$k+1$.
\end{enumerate}
The Firedrake implementation using Slate is similar to the scalar post-processing example
(see Listing \ref{lst:slateex-spp}); the element-wise linear systems \eqref{eq:fluxpp-a}--\eqref{eq:fluxpp-b}
can be expressed in UFL, and therefore the necessary Slate expressions to invert the local
systems follows naturally from the set of operations presented in Section \ref{subsec:slate-overview}.
We use the very sensitive parameter dependency in the post-processing methods to validate our software
implementation in \cite{zen-tabularasa}.

\section{Static condensation as a preconditioner}\label{sec:PCs}
Slate enables static condensation approaches to be expressed very concisely. Nonetheless, application
of a particular approach to different variational problems using Slate still requires a certain
amount of code repetition. By formulating each form of static condensation as a preconditioner,
code can be written once and then applied to any mathematically suitable problem.  Rather than
writing the static condensation by hand, in many cases, it is sufficient to just select the appropriate,
Slate-based, preconditioner.

For context, it is helpful to frame the problem in the particular context of the solver library.
Firedrake uses PETSc to provide linear solvers, and we implement our preconditioners as
PETSc \texttt{PC} objects.  These are defined to act on the problem residual, and return
a correction to the solution.  Specifically, we can think of (left) preconditioning the
matrix equation in residual form:
\begin{equation}\label{eq:linearsystem}
	r = r(A, b) \equiv b - Ax = 0
\end{equation}
by an operator $\boldsymbol{P}$ (which may not necessarily be linear) as a transformation
into an \emph{equivalent} system of the form
\begin{equation}\label{eq:pclinearsystem}
	\boldsymbol{P}r = \boldsymbol{P}(b - Ax) = 0.
\end{equation}
Given a current iterate $x_i$ the residual at the $i$-th iteration is simply $r_i \equiv b - A x_i$,
and $\boldsymbol{P}$ acts on the residual to produce an approximation to the error
$\epsilon_i \equiv x - x_i$. If $\boldsymbol{P}$ is an application of an exact
inverse, the residual is converted into an exact (up to numerical round-off) error.

We will denote the application of particular Krylov subspace method (KSP) for the
linear system \eqref{eq:linearsystem} as $\mathcal{K}_{x}(r(A, b))$. Upon preconditioning
the system via $\boldsymbol{P}$ as in \eqref{eq:pclinearsystem}, we write
\begin{equation}\label{eq:pcsolver}
	\mathcal{K}_{x}(\boldsymbol{P}r(A, b)).
\end{equation}
If \eqref{eq:pcsolver} is solved directly via the application of $A^{-1}$, then
$\boldsymbol{P}r(A, b) = A^{-1}b - x$. So, we have that
$\mathcal{K}_{x}(\boldsymbol{P}r(A, b)) = \mathcal{K}_{x}(r(I, A^{-1}b))$
produces the exact solution of \eqref{eq:linearsystem} in a single iteration of
$\mathcal{K}$. Having established notation, we now present our implementation
of static condensation via Slate by defining the appropriate operator, $\boldsymbol{P}$.

\subsection{Interfacing with PETSc via custom preconditioners}\label{subsec:slate+petsc}
The implementation of preconditioners for these systems requires manipulation not of
assembled matrices, but rather their symbolic representation.  To do this, we use the
preconditioning infrastructure developed by \citet{mitchelkirby}, which gives
preconditioners written in Python access to the symbolic problem description. In Firedrake,
this means all derived preconditioners have direct access to the UFL representation of
the PDE system. From this mathematical specification, we manipulate this appropriately via
Slate and provide operators assembled from Slate expressions to PETSc for further algebraic
preconditioning. Using this approach, we have developed a static condensation interface
for the hybridization of $\boldsymbol{H}(\text{div})\times L^2$ mixed problems, and a
generic interface for statically condensing hybridized systems. The advantage of writing
even the latter as a preconditioner is the ability to switch out the solution scheme for
the system, even when nested inside a larger set of coupled equations at \emph{runtime}.

\subsubsection{A static condensation interface for hybridization}\label{subsubsec:genhybridpc}
As discussed in sections \ref{subsec:hybrid-mixed-example} and \ref{subsubsec:HDG}, one
of the main advantages of using a hybridizable variant of a DG or mixed method is
that such systems permit the use of element-wise condensation and recovery. To facilitate
this, we provide a PETSc \texttt{PC} static condensation interface, \texttt{SCPC}.
This preconditioner takes the discretized system as in \eqref{eq:hybrid-sys}, and
performs the local elimination and recovery procedures. Slate expressions are
generated from the underlying UFL problem description.

More precisely, the incoming system has the form:
\begin{equation}\label{eq:hybrid-blocksystem}
	\begin{bmatrix}
		A_{e,e} & A_{e,c}\\
		A_{c,e} & A_{c,c}
	\end{bmatrix}
	\begin{Bmatrix}
		X_e \\
		X_c
	\end{Bmatrix} =
	\begin{Bmatrix}
		R_e \\ R_c
	\end{Bmatrix},
\end{equation}
\noindent where $X_e$ is the vector of unknowns to be eliminated, $X_c$ is the vector
of unknowns for the condensed field, and $R_e$, $R_c$ are the incoming right-hand sides.
The partitioning in \eqref{eq:hybrid-blocksystem} is determined by the \texttt{SCPC}
option: \texttt{pc\_sc\_eliminate\_fields}. Field indices are provided in the same way
one configures solver options to PETSc. These indices determine which
field(s) to statically condense into. For example, on a three-field
problem (with indices 0, 1, and 2), setting \texttt{-pc\_sc\_eliminate\_fields 0,1}
will configure \texttt{SCPC} to cell-wise eliminate field 0 and 1; the resulting
condensed system is associated with field 2.

In exact arithmetic, the \texttt{SCPC} preconditioner applies the inverse of the
Schur-complement factorization of \eqref{eq:hybrid-blocksystem}:
\begin{equation}\label{eq:hybrid-schur-PC}
	\boldsymbol{P} = 
	\begin{bmatrix}
		I & A_{e,e}^{-1}A_{e,c} \\
		0 & I
	\end{bmatrix}
	\begin{bmatrix}
		A_{e,e}^{-1} & 0\\
		0 & S^{-1}
	\end{bmatrix}
	\begin{bmatrix}
		I & 0 \\
		A_{c,e}A_{e,e}^{-1} & I
	\end{bmatrix},
\end{equation}
where $S = A_{c,c} - A_{c,e} A_{e,e}^{-1} A_{e,c}$ is the Schur-complement operator for the $X_c$ system.
The distinction here from block preconditioners via \texttt{fieldsplit} \citep{fieldsplit}, for example, is
that $\boldsymbol{P}$ does not require global actions; by design $A_{e,e}^{-1}$ can be inverted
locally and $S$ is sparse. As a result, $S$ can be assembled or applied exactly via Slate-generated
kernels.

In practice, the only globally coupled system requiring iterative inversion is $S$:
\begin{equation}\label{eq:traceksp}
	\mathcal{K}_{X_c}(\boldsymbol{P}_1 r(S, R_E)),
\end{equation}
where $R_E$ is the condensed right-hand side and $\boldsymbol{P}_1$
is another possible choice of preconditioner for $S$. Once $X_c$ is computed, $X_e$ is
reconstructed element-wise via inverting the local systems.

By construction, this preconditioner is suitable for both hybridized mixed and HDG
discretizations. It can also be used within other contexts, such as the static condensation
of continuous Galerkin discretizations \citep{guyan,irons} or primal-hybrid methods
\citep{cgprimalhybrid}. As with any PETSc preconditioner, solver
options can be specified for inverting $S$ via the appropriate options prefix (\texttt{condensed\_field}).
The resulting KSP for \eqref{eq:traceksp} is compatible with existing solvers and external
packages provided through the PETSc library. This allows users to experiment with a direct
method and then switch to a more parallel-efficient iterative solver
without changing the core application code.

\subsubsection{Preconditioning mixed methods via hybridization}\label{subsubsec:HybridPC}
The preconditioner \texttt{HybridizationPC} expands on the previous one, this time taking an
$\boldsymbol{H}(\text{div})\times L^2$ system and automatically forming the hybridizable
problem. This is accomplished through manipulating the UFL objects representing the discretized
PDE. This includes replacing argument spaces with their discontinuous counterparts, introducing
test functions on an appropriate trace space, and providing operators assembled from Slate
expressions in a similar manner as described in section \ref{subsubsec:genhybridpc}.

More precisely, let $AX = R$ be the incoming mixed saddle
point problem, where $R = \begin{Bmatrix}
R_{U} & R_P
\end{Bmatrix}^T$,
$X = \begin{Bmatrix}
U & P
\end{Bmatrix}^T$, and $U$
and $P$ are the velocity and scalar unknowns respectively. Then this preconditioner replaces $AX = R$ with the augmented
system:
\begin{equation}\label{eq:hybridmatsys}
	\begin{bmatrix}
		\widehat{A} & K^T\\
		K & 0
	\end{bmatrix}
	\begin{Bmatrix}
		\widehat{X}\\
		\Lambda
	\end{Bmatrix} =
	\begin{Bmatrix}
		\widehat{R}\\
		R_\Lambda
	\end{Bmatrix}
\end{equation}
\noindent where $\widehat{R} = \begin{Bmatrix}
\widehat{R}_U & R_P
\end{Bmatrix}^T$, $\widehat{R}_U$, $R_P$ are the right-hand sides for the flux
and scalar equations respectively, and $\widehat{\cdot}$ indicates modified
matrices and co-vectors with discontinuous functions. Here,
$\widehat{X} = \begin{Bmatrix}
U^d & P
\end{Bmatrix}^T$ are the hybridized (discontinuous) unknowns to be determined.

The preconditioning operator for the hybrid-mixed system \eqref{eq:hybridmatsys} has the form:
\begin{equation}\label{eq:PCHybridMixed}
\boldsymbol{\widehat{P}} =
	\begin{bmatrix}
		I & \widehat{A}^{-1}K^T\\
		0 & I
	\end{bmatrix}
	\begin{bmatrix}
		\widehat{A}^{-1} & 0\\
		0 & S^{-1}
	\end{bmatrix}
	\begin{bmatrix}
		I & 0\\
		K\widehat{A}^{-1} & I
	\end{bmatrix},
\end{equation}
\noindent where $S$ is the Schur-complement matrix $S = -K \widehat{A}^{-1} K^T$.
As before, a single globally coupled system for $\Lambda$ is required. The
recovery of $U^d$ and $P$ happens in the same manner as \texttt{SCPC}.

Since the flux is constructed in a discontinuous space $\boldsymbol{U}^d_h$, we must project the
computed solution into $\boldsymbol{U}_h \subset \boldsymbol{H}(\text{div})$. This can be done
cheaply via local facet averaging. The resulting solution is then updated via $U \leftarrow
\Pi_{\text{div}}U^d$, where $\Pi_{\text{div}}: \boldsymbol{U}_h^d \rightarrow \boldsymbol{U}_h$
is the projection mapping.
This ensures the residual for the original mixed problem is properly evaluated to
test for convergence. With $\boldsymbol{\widehat{P}}$ as in \eqref{eq:PCHybridMixed},
the preconditioning operator for the original $A X = R$ system is:
\begin{equation}\label{eq:PCMixed}
	\boldsymbol{P} =
	\Pi
	\boldsymbol{\widehat{P}}
	\Pi^T,
	\quad
	\Pi = \begin{bmatrix}
		\Pi_{\text{div}} & 0 & 0\\
		 0 & I & 0
	\end{bmatrix}.
\end{equation}

We note here that assembly of the right-hand for the $\Lambda$ system requires special attention. Firstly,
when Neumann conditions are present, then $R_\Lambda$ is not necessarily 0. Since the hybridization
preconditioner has access to the entire Python context (which includes a list of boundary conditions
and the spaces in which they are applied), surface integrals on the exterior boundary are added where
appropriate and incorporated in the generated Slate expressions. A more subtle issue that requires
extra care is the incoming right-hand side tested in $\boldsymbol{U}_h$.

The situation we are given is that we have $R_U = R_U(\boldsymbol{w})$ for $\boldsymbol{w} \in \boldsymbol{U}_h$,
but require $\widehat{R}_U(\boldsymbol{w}^d)$ for $\boldsymbol{w}^d \in \boldsymbol{U}^d_h$. For consistency,
we also require for any $\boldsymbol{w} \in \boldsymbol{U}_h$ that
\begin{equation}\label{eq:residual-consistency}
	\widehat{R}_U(\boldsymbol{w}) = R_U(\boldsymbol{w}).
\end{equation}
We can construct such a $\widehat{R}_U$ satisfying \eqref{eq:residual-consistency} in the following way.
By construction of the space $\boldsymbol{U}^d_h$, we have for
$\boldsymbol{\Psi}_i \in \boldsymbol{U}_h$:
\begin{equation}\label{eq:hdiv-basis-fct}
	\boldsymbol{\Psi}_i =
	\begin{cases}
	\boldsymbol{\Psi}_i^d & \boldsymbol{\Psi}_i \text{ associated with an exterior facet node},\\
	\boldsymbol{\Psi}_i^{d,+} + \boldsymbol{\Psi}_i^{d,-}
	& \boldsymbol{\Psi}_i \text{ associated with an interior facet node},\\
	\boldsymbol{\Psi}^d_i
	& \boldsymbol{\Psi}_i \text{ associated with a cell interior node},\\
	\end{cases}
\end{equation}
where $\boldsymbol{\Psi}^d_i, \boldsymbol{\Psi}_i^{d,\pm} \in \boldsymbol{U}^d_h$,
and $\boldsymbol{\Psi}_i^{d,\pm}$ are functions corresponding to the positive
and negative restrictions associated with the $i$-th facet node\footnote{
	These are the two ``broken" parts of $\boldsymbol{\Psi}_i$ on a particular
	facet connecting two elements. That is, for two adjacent
	cells, a basis function in $\boldsymbol{U}_h$ for a particular facet node
	can be decomposed into two basis functions in $\boldsymbol{U}^d_h$ defined on
	their respective sides of the facet.}.
We then define our ``broken'' right-hand side via the local definition:
\begin{equation}\label{eq:local-def-residual}
	\widehat{R}_U(\boldsymbol{\Psi}^d_i) =
	\frac{R_U(\boldsymbol{\Psi}_i)}{N_i},
\end{equation}
where $N_i$ is the number of cells that the degree of freedom corresponding to the basis function
$\boldsymbol{\Psi}_i \in \boldsymbol{U}_h$ touches. Using \eqref{eq:hdiv-basis-fct},
\eqref{eq:local-def-residual}, and the fact that $R_U$ is linear in its argument, we can verify that our
construction of $\widehat{R}_U$ satisfies \eqref{eq:residual-consistency}.

\section{Numerical studies}\label{sec:applications}
We now present results utilizing the Slate DSL and our static condensation
preconditioners for a set of test problems. Since we are using the interfaces outlined in
Section \ref{sec:PCs}, Slate is accessed indirectly and requires no manually-written solver
code for hybridization or static condensation/local recovery.
All parallel results were obtained on a
single fully-loaded compute node of dual-socket Intel E5-2630v4 (Xeon) processors
with $2 \times 10$ cores (2 threads per core) running at 2.2GHz. In order to avoid
potential memory effects due to the operating system migrating processes between sockets,
we pin MPI processes to cores.

Verification of the generated code is performed using parameter-sensitive convergence
tests. The study consists of running a variety of discretizations spanning the methods
outlined in Section \ref{subsec:examples}. Details and numerical results are made
public and can be viewed in \cite{zen-tabularasa} (see ``Code and data availability'').
All results are in full agreement with the theory.

\subsection{LDG-H method for a three-dimensional elliptic equation}\label{subsec:HDGCGcomp}
In this section, we take a closer look at the LDG-H method for the model elliptic
equation (sign-definite Helmholtz):
\begin{align}
	-\nabla\cdot\nabla p + p &= f \text{ in } \Omega = \lbrack 0, 1\rbrack^3,\label{eq:3delliptic}\\
	p &= g \text{ on } \partial\Omega,\label{eq:3delliptic-bc}
\end{align}
where $f$ and $g$ are chosen such that the analytic solution is
$p = \exp\lbrace\sin(\pi x)\sin(\pi y)\sin(\pi z)\rbrace$.
We use a regular mesh consisting $6 \cdot N^3$ tetrahedral elements
($N \in \lbrace 4, 8, 16, 32, 64 \rbrace$). First, we reformulate \eqref{eq:3delliptic}--\eqref{eq:3delliptic-bc}
as the mixed problem:
\begin{align}
	\boldsymbol{u} + \nabla p &= 0,\label{eq:3dmixed-a} \\
	\nabla\cdot\boldsymbol{u} + p &= f \label{eq:3dmixed-b}, \\
	p &= g \text{ on } \partial\Omega \label{eq:3dmixed-c}.
\end{align}
We start with linear polynomial approximations, up to cubic, for the LDG-H discretization
of \eqref{eq:3dmixed-a}--\eqref{eq:3dmixed-c}.
Additionally, we compute a post-processed scalar approximation $p_h^\star$ of the HDG solution.
This raises the approximation order of the computed solution by an additional degree.
In all numerical studies here, we set the HDG parameter $\tau = 1$. All results were computed
in parallel, utilizing a single compute node (described previously).

A continuous Galerkin (CG) discretization of the primal problem \eqref{eq:3delliptic}--\eqref{eq:3delliptic-bc}
serves as a reference for this experiment.
Due to the superconvergence in the post-processed solution for the HDG method, we
use CG discretizations of polynomial order 2, 3, and 4. This takes into account the
enhanced accuracy of the HDG solution, despite being initially computed as a lower-order
approximation. We therefore expect both methods to produce equally accurate solutions
to the model problem.

Our aim here is not to compare the performance of HDG and CG, which has been investigated
elsewhere (for example, see \citet{kirbyEtAl, yakovlev2016cg}). Instead, we provide a
reference that the reader might be more familiar with in order to evaluate whether our
software framework produces a sufficiently performant HDG implementation relative to
what might be expected.

For the CG discretization, we use a matrix-explicit iterative solver
consisting of the conjugate gradient method preconditioned with
hypre's boomerAMG implementation of algebraic multigrid
(AMG) \cite{falgout2006design}.  We use the preconditioner described
in Section \ref{subsubsec:genhybridpc} to statically condense the
LDG-H system, using the same iterative solver as the CG method for the
Lagrange multipliers. This is sensible, as the global trace operator
defines a symmetric positive-definite operator. To avoid over-solving,
we iterate to a relative tolerance such that the discretization error is
minimal for a given mesh.

\subsubsection{Error versus execution time}
\begin{figure}[htbp]
	\centering
	\begin{subfigure}{.5\textwidth}
		\centering
		\captionsetup{width=.9\linewidth}
		\includegraphics[width=0.85\textwidth]{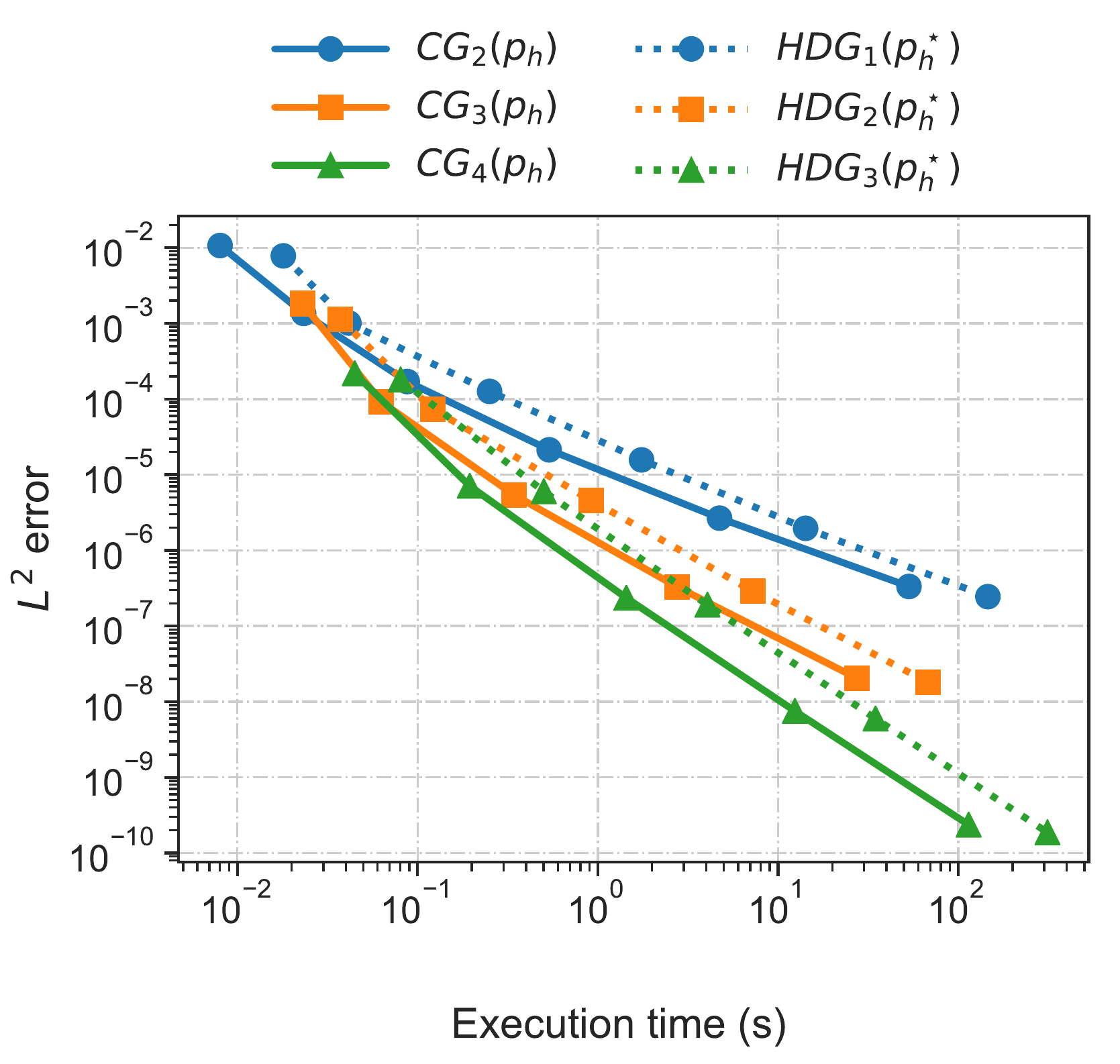}
		\caption{Error against execution time for the CG and HDG
			with post-processing ($\tau = 1$) methods.}
		\label{fig:cg-hdg-budget}
	\end{subfigure}%
	\begin{subfigure}{.5\textwidth}
		\centering
		\captionsetup{width=.9\linewidth}
		\includegraphics[width=0.825\textwidth]{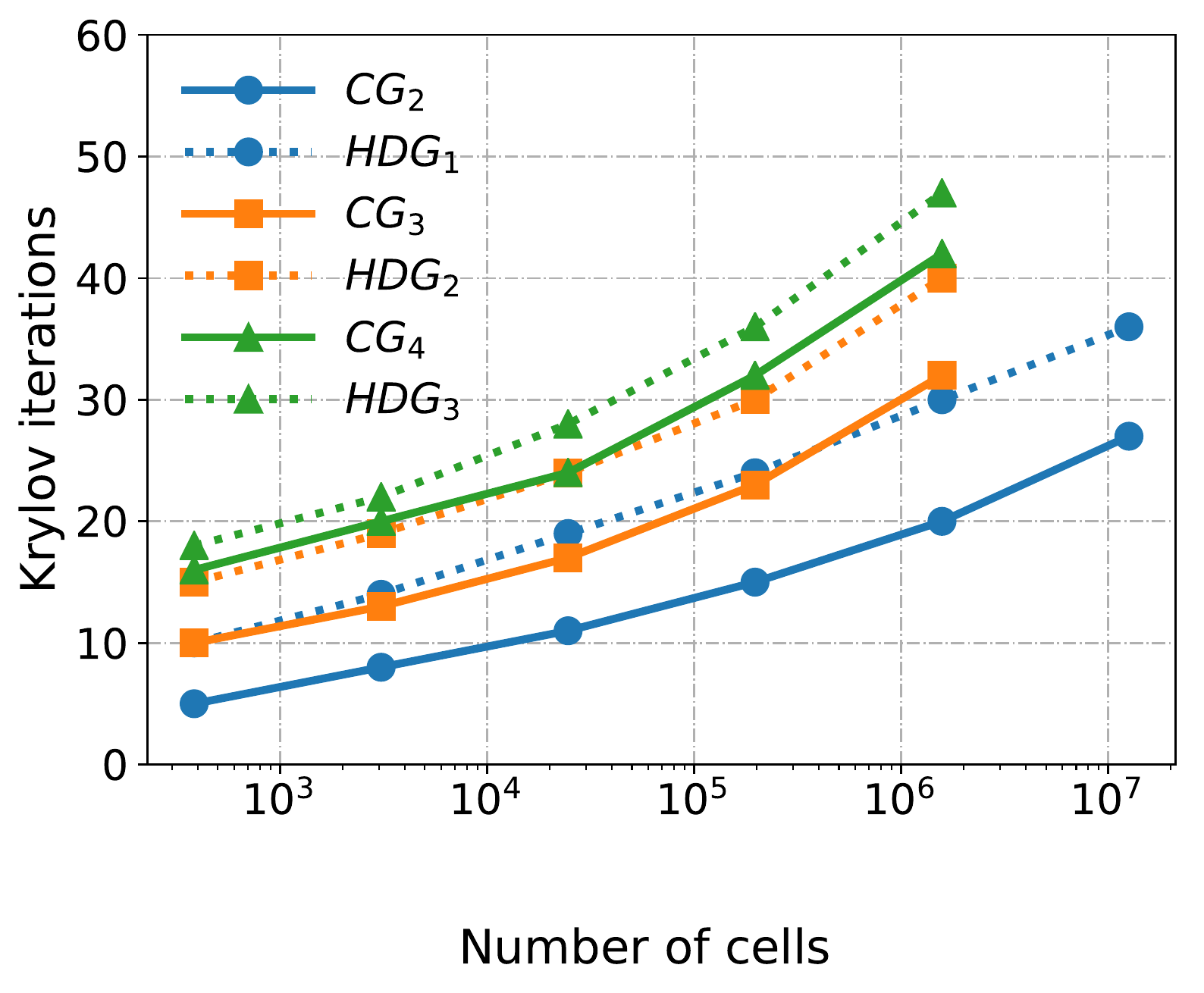}
		\caption{Krylov iterations of the AMG-preconditioned conjugate gradient algorithm
			(to reach discretization error) against number of cells.}
		\label{fig:cg-hdg-data}
	\end{subfigure}
	\caption{Comparison of continuous Galerkin and LDG-H solvers for the model three-dimensional
		positive-definite Helmholtz equation.}
	\label{fig:cgvshdg}
\end{figure}

The total execution time is recorded for the CG and HDG solvers,
which includes the setup time for the AMG preconditioner, matrix-assembly, and
the time-to-solution for the Krylov method. In the HDG case, we include the
time spent building the Schur-complement for the traces, local recovery of
the scalar and flux approximations, and post-processing. The $L^2$-error against
execution time is summarized in Figure \ref{fig:cg-hdg-budget}.

The HDG method of order $k-1$ ($HDG_{k-1}$) with post-processing,
as expected, produces a solution which is as accurate as the CG method of order $k$ ($CG_k$).
While the full HDG system is never explicitly assembled, the larger execution time is a result
of several factors. The total number of trace unknowns for the $HDG_1$, $HDG_2$, and $HDG_3$
discretizations is roughly four, three, and two times larger (resp.) than the corresponding number
of CG unknowns. Therefore, each iteration is more expensive. Moreover, we also observe that the
trace system requires more Krylov iterations to reach discretization error. The
gap in total number of iterations starts to close as the approximation degree increases
(see Figure \ref{fig:cg-hdg-data}).
The extra cost of HDG due to the larger degree-of-freedom count and the need to perform
local tensor inversion is offset by the local conservation and stabilization properties
which are useful for fluid dynamics applications.

\subsubsection{Break down of solver time}\label{subsubsec:hdgbreakdown}
The HDG method requires many more degrees of freedom than CG or
primal DG methods. This is largely due to the fact that the HDG method simultaneously
approximates the primal solution and its velocity. The global matrix for the traces is larger than
the one for the CG system at low polynomial order. The execution time
for HDG is then compounded by a more expensive global solve. We remind the reader that our
goal here is to verify that the solve time for HDG is as might be expected given the problem size.
\begin{figure}[htbp]
	\centering
	\includegraphics[width=0.6\textwidth]{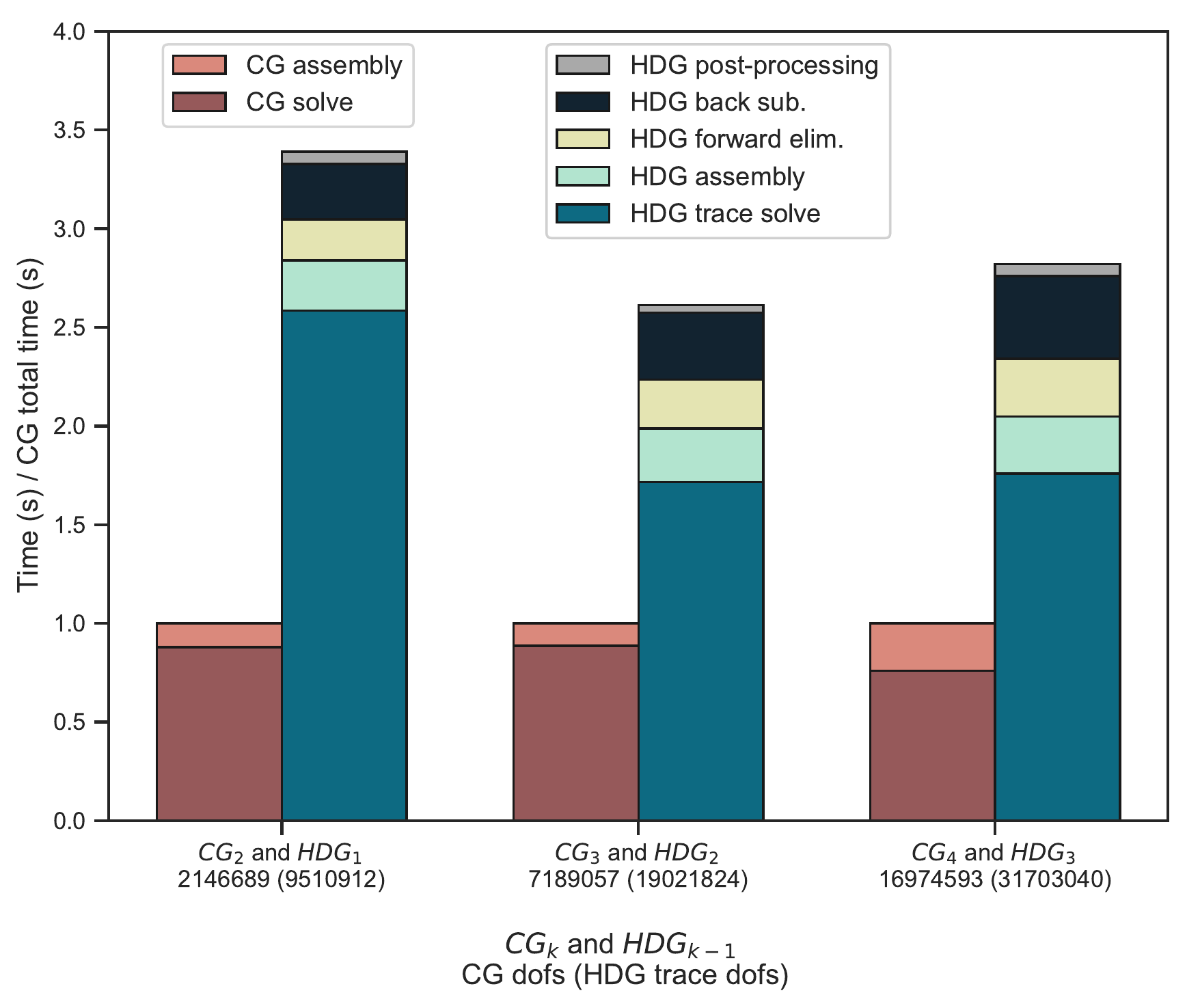}
	\caption{Break down of the $CG_k$ and $HDG_{k-1}$ execution times on a
		$6 \cdot 64^3$ simplicial mesh.}
	\label{fig:cg-hdg-timings}
\end{figure}
\begin{table}[tbhp]
	\captionsetup{position=top}
	\caption{Breakdown of the raw timings for the $HDG_{k-1}$ ($\tau = 1$) and $CG_k$ methods,
		$k = 2$, $3$, and $4$. Each method corresponds to a mesh size $N = 64$ on a fully-loaded
		compute node.}
	\label{tab:hdg-breakdown}
	\centering
	\begin{tabular}{lcccccc}
	\hline
	\multirow{2}{*}{Stage}
	& \multicolumn{2}{c}{$HDG_1$}
	& \multicolumn{2}{c}{$HDG_2$}
	& \multicolumn{2}{c}{$HDG_3$}
	\\
	& $t_{\text{stage}}$ (s) & \% $t_{\text{total}}$
	& $t_{\text{stage}}$ (s) & \% $t_{\text{total}}$
	& $t_{\text{stage}}$ (s) & \% $t_{\text{total}}$ \\ \hline
	Matrix assembly (static cond.) &  1.05 &  7.49 \% &  6.95 &  10.40 \% &  31.66 &  10.27 \% \\
	Forward elimination &  0.86 &  6.13 \% &  6.32 &  9.45 \% &  31.98 &  10.37 \% \\
	Trace solve &  10.66 &  76.24 \% &  43.89 &  65.66 \% &  192.31 &  62.36 \% \\
	Back substitution &  1.16 &  8.28 \% &  8.71 &  13.03 \% &  45.81 &  14.85 \% \\
	Post processing &  0.26 &  1.86 \% &  0.98 &  1.46 \% &  6.62 &  2.15 \% \\
	\hline
	HDG Total & 13.98 & & 66.85 & & 308.37 & \\ \hline
	& \multicolumn{2}{c}{$CG_2$}
	& \multicolumn{2}{c}{$CG_3$}
	& \multicolumn{2}{c}{$CG_4$}
	\\
	& $t_{\text{stage}}$ (s) & \% $t_{\text{total}}$
	& $t_{\text{stage}}$ (s) & \% $t_{\text{total}}$
	& $t_{\text{stage}}$ (s) & \% $t_{\text{total}}$ \\ \hline
	Matrix assembly (monolithic) &  0.50 &  12.01 \% &  2.91 &  11.39 \% &  26.37 &  24.11 \% \\
	Solve &  3.63 &  87.99 \% &  22.67 &  88.61 \% &  82.99 &  75.89 \% \\
	\hline
	CG Total & 4.12 & & 25.59 & & 109.36 & \\ \hline
	\end{tabular}
\end{table}

Figure \ref{fig:cg-hdg-timings} displays the execution times on a simplicial mesh consisting of
$1.5$ million elements. The execution times have been normalized by the CG total time, showing
that the HDG method is roughly 3 times the execution time of the CG method. This is
expected given the larger degree-of-freedom count. The raw numerical
breakdown of the HDG and CG solvers are shown in Table \ref{tab:hdg-breakdown}. We isolate
each component of the HDG method contributing to the total execution time.
Local operations include static condensation (trace operator assembly), forward elimination
(right-hand side assembly for the trace system), backwards substitution to recover
the scalar and velocity unknowns, and local post-processing of the primal solution. For all $k$,
our HDG implementation is solver-dominated.

Both operator and right-hand side assembly are dominated by the costs of inverting a local
square mixed matrix coupling the primal and dual variables, which is performed directly
via an LU factorization. They should therefore be of the same magnitude
in time spent. We observe that this is the case across all degrees (ranging between
approximately 6---11\% of total execution time for both local operations).
Back-substitution takes roughly the same time as the static condensation and forward elimination
stages (between 8---15\% of execution time across all $k$).
This is expected, as these operations are all dominated by the cost of inverting the
local matrix coupling the scalar and velocity degrees of freedom. The slight increase in time
is due splitting the local equations into two local solvers: one for $p_h$ and another for
the velocity. Finally, the additional cost of post-processing accrues negligible time
(roughly 2\% of execution time across all degrees).

We note that caching of local tensors does not occur. Each pass to perform the
local eliminations and backwards reconstructions rebuilds the local element tensors.
It is not clear at this time whether the performance gained from avoiding rebuilding
the local operators will offset the memory costs of storing the local matrices. Moreover,
in time-dependent problems where the operators may contain state-dependent variables,
rebuilding local matrices will be necessary in each time-step regardless.

\subsection{Hybridized-mixed solver for the shallow water equations}\label{subsec:williamson5}
A primary motivator for our interest in hybridized methods revolves around developing efficient
solvers for problems in geophysical flows. In section, we present some results
integrating the nonlinear shallow water equations on the sphere using test case 5 (flow past a mountain)
from \citet{williamson}. We use the framework of compatible finite elements
\citep{cottershipton,cotterthuburn}.

\subsubsection{Semi-implicit finite element discretization}
We start with the vector-invariant rotating nonlinear shallow water system defined on
a two-dimensional spherical surface $\Omega$ embedded in $\mathbb{R}^3$:
\begin{align}
	\frac{\partial \boldsymbol{u}}{\partial t} +
	\left(\nabla^\perp \cdot\boldsymbol{u} + f\right)\boldsymbol{u}^\perp
	+ \nabla\left(g\left(D + b\right) + \frac{1}{2}|\boldsymbol{u}|^2\right)
	&= 0,
	\label{eq:nswe-a}\\
	\frac{\partial D}{\partial t} + \nabla\cdot\left(\boldsymbol{u}D\right)
	&= 0,
	\label{eq:nswe-b}
\end{align}
\noindent where $\boldsymbol{u}$ is the fluid velocity, $D$ is the depth field,
$f$ is the Coriolis parameter, $g$ is the acceleration due to gravity,
$b$ is the bottom topography, and $(\cdot)^\perp \equiv \zhat \times \cdot$,
with $\zhat$ being the unit normal to the surface
$\Omega$.

After discretizing in space and time using a semi-implicit scheme and Picard linearization, following
\citet{natale2017variational} and based on \citet{melvin2018mixed} (see Appendix \ref{app:semiimplicit} for a
summary of the entire solution strategy), we must solve an indefinite saddle point system at each step:
\begin{equation}\label{eq:SPSSWE}
	\begin{bmatrix}
		A & B\\
		C & D
	\end{bmatrix}
	\begin{Bmatrix}
		\Delta U\\
		\Delta D
	\end{Bmatrix} =
	\begin{Bmatrix}
		R_{\boldsymbol{u}}\\
		R_D
	\end{Bmatrix}.
\end{equation}
The approximations $\Delta U$ and $\Delta D$ are sought in the mixed finite element spaces
$\boldsymbol{U}_h \subset \boldsymbol{H}(\text{div})$
and $V_h \subset L^2$ respectively. Pairings including the RT or BDM spaces such as $\text{RT}_k \times \text{DG}_{k-1}$
or $\text{BDM}_k \times \text{DG}_{k-1}$ fall within the set of compatible mixed spaces ideal for geophysical fluids
\citep{cottershipton, natale2016compatible, melvin2018mixed}. In particular, the lowest-order RT
method ($\text{RT}_1 \times \text{DG}_0$) on a structured quadrilateral grid (such as the latitude-longitude grid
used many operational dynamical cores) corresponds to the Arakawa C-grid finite difference discretization.

In staggered finite difference models, the standard approach for solving \eqref{eq:SPSSWE} is to neglect the
Coriolis term and eliminate the velocity unknowns $\Delta U$ to obtain a discrete elliptic problem for
which smoothers like Richardson iterations or relaxation methods are convergent. This is more problematic in the compatible
finite element framework, since $A$ has a dense inverse. Instead, we use the preconditioner described in
Section \ref{subsubsec:HybridPC} to form the hybridized problem and eliminate both $\Delta U$ and
$\Delta D$ locally.

\subsubsection{Atmospheric flow over a mountain}\label{subsubsec:swemtn}
As a test problem, we solve test case 5 of \citet{williamson}, on the surface of an Earth-sized sphere.
We refer the reader to \citet{cottershipton, shipton2018higher} for a more comprehensive
study on mixed finite elements for shallow water systems of this type.
We use the mixed finite element pairs $(\text{RT}_1, \text{DG}_0)$ (lowest-order RT method)
and $(\text{BDM}_2, \text{DG}_1)$ (next-to-lowest order BDM method) for the velocity and depth spaces.
The sphere mesh is generated from 7 refinements of an icosahedron, resulting in a triangulation
consisting of 327,680 elements in total. The discretization information is summarized in Table \ref{tab:grid}.
\begin{table}[!tbhp]
	\captionsetup{position=top}
	\caption{The number of unknowns to be determined are
	summarized for each compatible finite element method. Resolution is the same
	for both methods.}
	\label{tab:grid}
	\centering

\begin{tabular}{ccccccc}
\hline
\multicolumn{7}{c}{\textbf{Discretization properties}} \\
\multicolumn{2}{c}{\multirow{2}{*}{Mixed method}} & \multirow{2}{*}{\# cells} &
\multirow{2}{*}{$\Delta x$} & Velocity & Depth & \multirow{2}{*}{Total (millions)} \\
& & & &  unknowns & unknowns & \\ \hline
\multicolumn{2}{c}{$\text{RT}_1 \times \text{DG}_0$} &
\multirow{2}{*}{327,680} &
\multirow{2}{*}{$\approx 43$ km} & 491,520 & 327,680 & 0.8 M \\
\multicolumn{2}{c}{$\text{BDM}_2 \times \text{DG}_1$} & & & 2,457,600
& 983,040 & 3.4 M \\
\hline
\end{tabular}

\end{table}

We run for a total of 25 time-steps, with a fixed number of 4 Picard iterations
in each time-step. We compare the overall simulation time using two different solver
configurations for the implicit linear system. First, we use a flexible variant of GMRES
\footnote{We use a flexible version of GMRES on the outer system since we use an
additional Krylov solver to iteratively invert the Schur-complement.}
acting on the outer system with an approximate Schur complement preconditioner:
\begin{equation}\label{eq:approx-sc-factorization}
	\boldsymbol{P}_{\text{SC}} =
	\begin{bmatrix}
		I & 0 \\
		CA^{-1} & I
	\end{bmatrix}
	\begin{bmatrix}
		A^{-1} & 0 \\
		0 & \widetilde{S}^{-1}
	\end{bmatrix}
	\begin{bmatrix}
		I & -A^{-1}B \\
		0 & I
	\end{bmatrix},
\end{equation}
where $\widetilde{S} = D - C \text{diag}(A)^{-1} B$, and $\text{diag}(A)$ is a diagonal
approximation to the velocity mass matrix.
The Schur-complement system is inverted via GMRES due to the asymmetry from the
Coriolis term, with the inverse of $\widetilde{S}$
as the preconditioning operator. The sparse approximation $\widetilde{S}$ is
inverted using PETSc's smoothed aggregation multigrid (GAMG).
The Krylov method is set to terminate once the preconditioned residual norm is
reduced by a factor of $10^{8}$. $A^{-1}$ is computed approximately using a
single application of incomplete LU (zero fill-in).

Next, we use only the application of our hybridization preconditioner (no outer iterations),
which replaces the original linearized mixed system with its hybrid-mixed equivalent. After
hybridization, we have the problem: find
$(\Delta\boldsymbol{u}^{d}_h, \Delta D_h, \lambda_h)
\in \boldsymbol{U}^{d}_h\times V_h\times M_h$ such that
\begin{align}
	\cellinner{\boldsymbol{w}, \Delta\boldsymbol{u}^d_h}_{\mathcal{T}_h}
	+ \frac{\Delta t}{2}\left(\boldsymbol{w}, f\left(\Delta\boldsymbol{u}^{d}_h\right)^{\perp}\right)_{\mathcal{T}_h}
	- \frac{\Delta t}{2}\cellinner{\nabla\cdot\boldsymbol{w}, g\Delta D_h}_{\mathcal{T}_h}
	+ \facetinner{\jump{\boldsymbol{w}}, \lambda_h}_{\partial\mathcal{T}_h}
	&= \widehat{R}_{\boldsymbol{u}}, &\forall \boldsymbol{w} \in \boldsymbol{U}^{d}_h,
	\label{eq:hybrid-mixedswe-a}\\
	\cellinner{\phi, \Delta D_h}_{\mathcal{T}_h}
	+ \frac{\Delta t}{2}\cellinner{\phi, H\nabla\cdot\Delta\boldsymbol{u}^{d}_h}_{\mathcal{T}_h}
	&= R_{D}, &\forall \phi \in V_h,
	\label{eq:hybrid-mixedswe-b}\\
	\facetinner{\gamma, \jump{\Delta\boldsymbol{u}^{d}_h}}_{\partial\mathcal{T}_h}
	&= 0, &\forall \gamma \in M_h.
	\label{eq:hybrid-mixedswe-c}
\end{align}
Note that the space $M_h$ is chosen such that these trace functions
when restricted to a facet $e \in \partial\mathcal{T}_h$ are from the same polynomial
space as $\Delta\boldsymbol{u}_h\cdot \boldsymbol{n}$ restricted to that same facet.
Additionally, it can be shown that $\lambda_h$ is an approximation to
$\Delta t g\Delta D/2$.

The resulting three-field problem for \eqref{eq:hybrid-mixedswe-a}--\eqref{eq:hybrid-mixedswe-c}
has the matrix form:
\begin{equation}\label{eq:supphybridswe}
	\begin{bmatrix}
		\widehat{A} & K^T \\
		K & 0
	\end{bmatrix}
	\begin{Bmatrix}
		\Delta X\\
		\Lambda
	\end{Bmatrix}
	=
	\begin{Bmatrix}
		\widehat{R}_{\Delta X}\\
		0
	\end{Bmatrix}
\end{equation}
\noindent where $\widehat{A}$ is the discontinuous operator coupling
$\Delta X = \begin{Bmatrix}
\Delta U^d & \Delta D
\end{Bmatrix}^T$, and $R_{\Delta X} = \begin{Bmatrix}
\widehat{R}_{\boldsymbol{u}} & R_{D}
\end{Bmatrix}^T$ are the problem residuals. An exact Schur-complement factorization is
performed on \eqref{eq:supphybridswe}, using Slate to generate the local elimination kernels.
We use the same set of solver options for the inversion of $\widetilde{S}$ in
\eqref{eq:approx-sc-factorization} to invert the Lagrange multiplier system. The
increments $\Delta U^d$ and $\Delta D$ are recovered
locally, using Slate-generated kernels. Once recovery is complete,
$\Delta U^d$ is injected back into the conforming
$\boldsymbol{H}(\text{div})$ finite element space via $\Delta U \leftarrow \Pi_{\text{div}}\Delta U^d$.
Based on the discussion in section \ref{subsubsec:HybridPC}, we apply:
\begin{equation}
	\boldsymbol{P}_{\text{hybrid}} = \Pi \left(
	\begin{bmatrix}
		I & \widehat{A}^{-1}K^T\\
		0 & I
	\end{bmatrix}
	\begin{bmatrix}
		\widehat{A}^{-1} & 0\\
		0 & S^{-1}
	\end{bmatrix}
	\begin{bmatrix}
		I & 0\\
		K\widehat{A}^{-1} & I
	\end{bmatrix}
	\right) \Pi^T.
\end{equation}
\begin{table}[!tbhp]
	\captionsetup{position=top}
	\caption{Preconditioner solve times for a 25-step run with $\Delta t = 100$s. These are
		cumulative times in each stage of the two preconditioners throughout the entire
		profile run. We display the average iteration count (rounded to the nearest integer)
		for both the outer and the inner Krylov solvers. The significant speedup when using
		hybridization is a direct result of eliminating the outer-most solver.}
	\label{tab:petsc-profile}
	\centering
\begin{tabular}{cccccc}\hline
\multicolumn{6}{c}{\textbf{Preconditioner and solver details}} \\
\multirow{2}{*}{Mixed method} &
\multirow{2}{*}{Preconditioner} &
\multirow{2}{*}{$t_{\text{total}}$ (s)} &
Avg. outer & Avg. inner &
\multirow{2}{*}{$\frac{t_{\text{total}}^{\text{SC}}}{t_{\text{total}}^{\text{hybrid.}}}$}\\
& & & its. & its. & \\ \hline
\multirow{2}{*}{$\text{RT}_1 \times \text{DG}_0$}& approx. Schur.
($\boldsymbol{P}_{\text{SC}}$)&  15.137 & 2 & 8 &

\multirow{2}{*}{3.413} \\& hybridization
($\boldsymbol{P}_{\text{hybrid}}$) &  4.434 & None & 2 &
\\
\hline\multirow{2}{*}{$\text{BDM}_2 \times \text{DG}_1$}& approx. Schur.
($\boldsymbol{P}_{\text{SC}}$) &  300.101 & 4 & 9 &

\multirow{2}{*}{5.556} \\& hybridization
($\boldsymbol{P}_{\text{hybrid}}$) &  54.013 & None & 6 &
\\
\hline
\end{tabular}
\end{table}

\begin{table}[!tbhp]
	\captionsetup{position=top}
	\caption{Breakdown of the cost (average) of a single application of the
	preconditioned flexible GMRES algorithm and hybridization. 
	Hybridization takes approximately the same time per iteration.}
	\label{tab:solver-breakdown}
	\centering

\begin{tabular}{llcccc}
\hline
\multirow{2}{*}{Preconditioner} &
\multirow{2}{*}{Stage} &
\multicolumn{2}{c}{$\text{RT}_1 \times \text{DG}_0$} &
\multicolumn{2}{c}{$\text{BDM}_2 \times \text{DG}_1$} \\
& & $t_{\text{stage}}$ (s) & \% $t_{\text{total}}$
& $t_{\text{stage}}$ (s) & \% $t_{\text{total}}$ \\ \hline

\multirow{4}{*}{approx. Schur ($\boldsymbol{P}_{\text{SC}}$)}
& Schur solve &  0.07592 &  91.28 \% &  0.78405 &  93.53 \% \\
& invert velocity mass: $A$ &  0.00032 &  0.39 \% &  0.00678 &  0.81 \% \\
& apply inverse: $A^{-1}$ &  0.00041 &  0.49 \% &  0.00703 &  0.84 \% \\
& gmres other &  0.00652 &  7.84 \% &  0.04041 &  4.82 \% \\
\hline
& Total & 0.08317 & & 0.83827 & \\ \hline
\multirow{5}{*}{hybridization ($\boldsymbol{P}_{\text{hybrid}}$)}
& Transfer: $R_{\Delta X}\rightarrow\widehat{R}_{\Delta X}$ &  0.00322 &  7.26 \% &  0.00597 &  1.10 \% \\
& Forward elim.: $-K\widehat{A}^{-1}\widehat{R}_{\Delta X}$ &  0.00561 &  12.64 \% &  0.12308 &  22.79 \% \\
& Trace solve &  0.02289 &  51.63 \% &  0.28336 &  52.46 \% \\
& Back sub. &  0.00986 &  22.23 \% &  0.12220 &  22.62 \% \\
& Projection: $\Pi_{\text{div}}\Delta\widehat{U}$ &  0.00264 &  5.96 \% &  0.00516 &  0.96 \% \\
\hline
& Total & 0.04434 & & 0.54013 & \\ \hline
\end{tabular}

\end{table}

Table \ref{tab:petsc-profile} displays a summary of our findings. When using hybridization,
we observe a significant reduction in time spent during the implicit solve stage compared
to the approximate Schur-complement approach. This is primarily because we reduce
the number of required ``outer'' iterations to zero; the hybridization preconditioner
is performing an \emph{exact} factorization of the global hybridized operator.
This is empirically supported when considering the per-iteration solve times, summarized in
Table \ref{tab:solver-breakdown}. Hybridization and the approximate Schur complement
preconditioner are comparable in terms of average execution
time, with hybridization being slightly faster per application. This further
demonstrates that the primary cause for the longer execution time of the latter is a
direct result of the additional outer iterations induced from using an approximate
factorization to achive the same reduction in the overall residual.

\begin{figure}[!ht]
  \begin{minipage}[b]{0.33\linewidth}
    \centering
    \includegraphics[width=.95\linewidth]{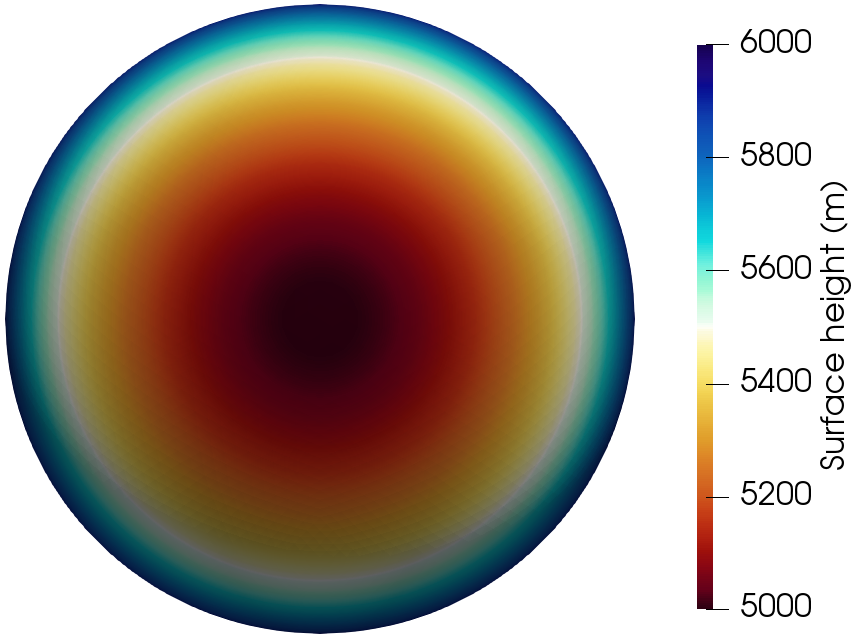} 
    \subcaption{Initial condition (day 0)} 
    \vspace{4ex}
  \end{minipage}%%
  \begin{minipage}[b]{0.33\linewidth}
    \centering
    \includegraphics[width=.95\linewidth]{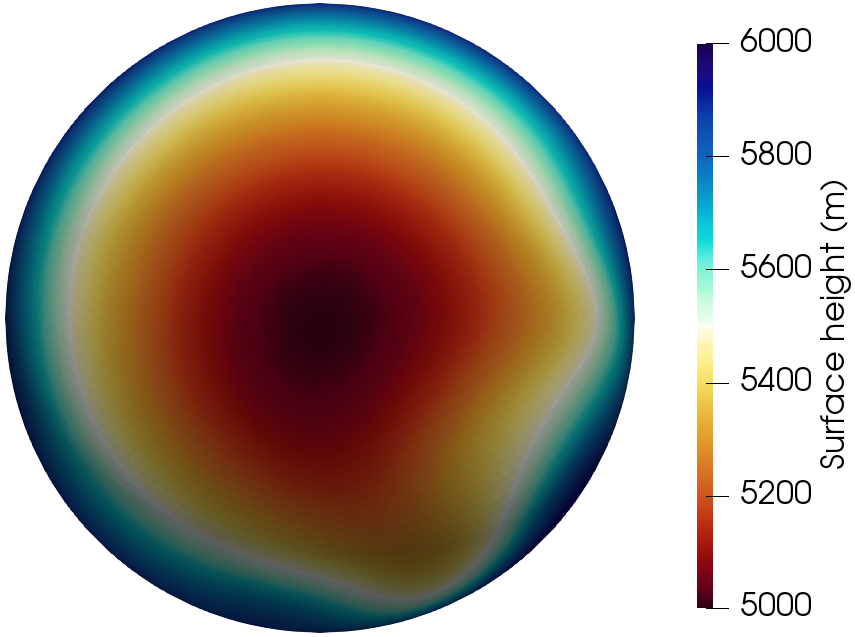} 
    \subcaption{Day 5} 
    \vspace{4ex}
  \end{minipage}
  \begin{minipage}[b]{0.33\linewidth}
    \centering
    \includegraphics[width=.95\linewidth]{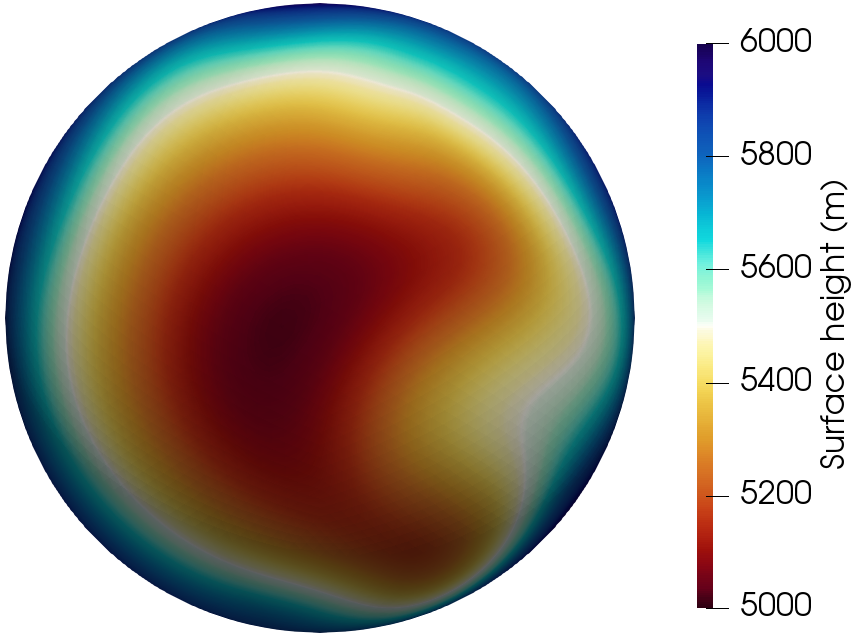} 
    \subcaption{Day 10} 
    \vspace{4ex}
  \end{minipage}%% 
  \begin{minipage}[b]{0.33\linewidth}
    \centering
    \includegraphics[width=.95\linewidth]{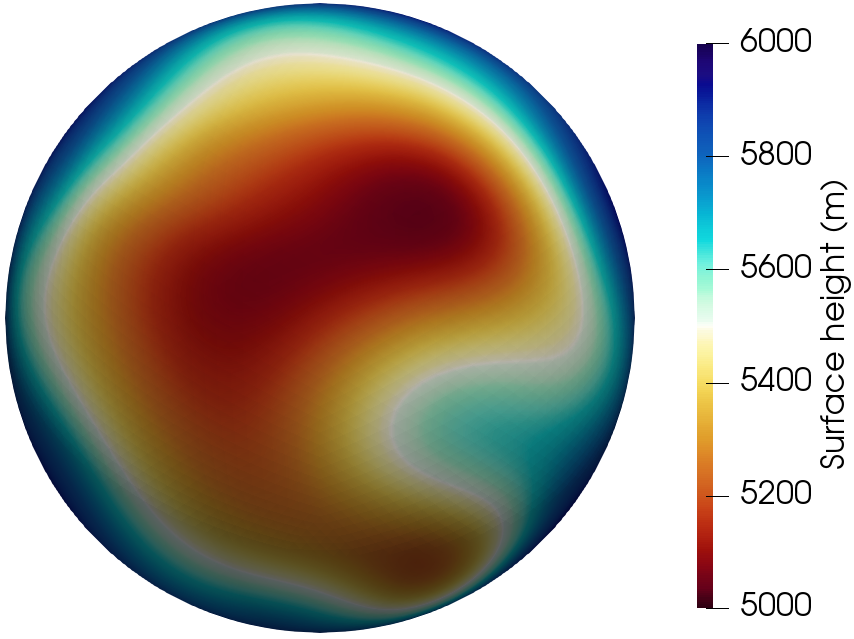} 
    \subcaption{Day 15} 
    \vspace{4ex}
  \end{minipage} 
  \caption{Snapshots (view from the northern pole) from the isolated mountain
		test case. The surface height (m) at days 5, 10, and 15.
		The snapshots were generated on a mesh with $20, 480$ simplicial cells, a
		$\text{BDM}_2 \times \text{DG}_1$ discretization, and $\Delta t = 500$ seconds.
		The linear system during each Picard cycle was solved using the hybridization preconditioner.}
		\label{fig:williamson-shots}
\end{figure}

We also measure the reductions in the true-residual of the linear system
\eqref{eq:SPSSWE}. Our hybridized method reduces the residual by a factor of
$10^{8}$ on average, which coincides with the specified relative tolerance for
the Krylov method on the trace system. Snapshots of a (coarser) 15 day simulation
are provided in Figure \ref{fig:williamson-shots} using the semi-implicit scheme
described in this paper. We refer the reader to \citet{shipton2018higher}
for an exposition of shallow water test cases featuring the use of a hybridized implicit solver
(as described in Section \ref{subsubsec:HybridPC}).

\subsection{Rotating linear Boussinesq model}\label{subsec:GW}
As a final example, we consider the simplified atmospheric model obtained
from a linearization of the compressible Boussinesq equations in a
rotating domain:
\begin{align}
	\frac{\partial \boldsymbol{u}}{\partial t} + 2\boldsymbol{\Omega}\times\boldsymbol{u}
	&= \nabla p + b\hat{\boldsymbol{z}}, \label{eq:gw_a}\\
	\frac{\partial p}{\partial t} &= -c^2 \nabla\cdot\boldsymbol{u},\label{eq:gw_b}\\
	\frac{\partial b}{\partial t} &= -N^2\boldsymbol{u}\cdot\hat{\boldsymbol{z}},\label{eq:gw_c}
\end{align}
where $\boldsymbol{u}$ is the fluid velocity, $p$ the pressure, $b$ is the buoyancy, $\boldsymbol{\Omega}$
the planetary angular rotation vector, $c$ is the speed of sound ($\approx 343 \text{m}\text{s}^{-1}$),
and $N$ is the buoyancy frequency ($\approx 0.01\text{s}^{-1}$). Equations \eqref{eq:gw_a}--\eqref{eq:gw_c}
permit fast-moving acoustic waves driven by perturbations in $b$. This
is the same model presented in \citet{skamarock1994efficiency}, which uses a quadratic equation of state to avoid some of the complications of the full compressible Euler equations (the hybridization of which we shall address in future work).
We solve these equations subject to the
rigid-lid condition $\boldsymbol{u}\cdot\boldsymbol{n} = 0$ on all boundaries.

Our domain consists of a spherical annulus, with the mesh constructed from a
horizontal ``base" mesh of the surface of a sphere of radius $R$, extruded
upwards by a height $H_{\Omega}$. The vertical
discretization is a structured one-dimensional grid, which facilitates the
staggering of thermodynamic variables, such as $b$. We consider two
kinds of meshes: one obtained by extruding an icosahedral
sphere mesh, and another from a cubed sphere.

Since our mesh has a natural tensor-product structure, we construct suitable finite element spaces
constructed by taking the tensor product of a horizontal space with a vertical space. To ensure
our discretization is ``compatible,'' we use the one- and two-dimensional de-Rham complexes:
$V_0 \overset{\partial_z}{\rightarrow}V_1$ and $U_0 \overset{\nabla^{\perp}}{\rightarrow}
U_1 \overset{\nabla\cdot}{\rightarrow}U_2$. We can then construct the three-dimensional complex:
$W_0 \overset{\nabla}{\rightarrow} W_1 \overset{\nabla\times}{\rightarrow} W_2\overset{\nabla\cdot}{\rightarrow} W_3$,
where
\begin{align}
	W_0 &= U_0\otimes V_0, \label{eq:complex-a}\\
	W_1 &= \Hcurl(U_1\otimes V_0)\oplus
	\Hcurl(U_0\otimes V_1) = W_1^v \oplus W_1^h,\label{eq:complex-b}\\
	W_2 &= \Hdiv(U_2\otimes V_0)\oplus
	\Hdiv(U_1\otimes V_1) = W_2^v \oplus W_2^h,\label{eq:complex-c}\\
	W_3 &= U_2\otimes V_1.\label{eq:complex-d}
\end{align}
Here, $\Hcurl$ and $\Hdiv$ denote operators which ensure the correct Piola tranformations are
applied when mapping from physical to reference element. We refer the reader to \citep{mcraeEtAl}
for an overview of constructing tensor-product finite element spaces. For the analysis of compatible
finite element discretizations and their relation to the complex \eqref{eq:complex-a}--\eqref{eq:complex-d},
we refer the reader to \cite{natale2016compatible,cotterthuburn,cottershipton}.
Each discretization used in this section is constructed from more familiar finite element families, shown in
Table \ref{tab:3dmixedmethods}.
\begin{table}[!htb]
	\captionsetup{position=top}
	\caption{Vertical and horizontal spaces for the three-dimensional compatible finite element discretization
		of the linear Boussinesq model. The $\text{RT}_k$ and $\text{BDFM}_{k+1}$ methods are
		constructed on triangular prism elements, while the $\text{RTCF}_k$ method is defined on
		extruded quadrilateral elements.}
	\label{tab:3dmixedmethods}
	\centering
	\begin{tabular}{cccccc}\hline
		\multicolumn{6}{c}{\textbf{Compatible finite element spaces}} \\
		Mixed method & $V_0$ & $V_1$ & $U_0$ & $U_1$ & $U_2$ \\ \hline
		$\text{RT}_k$ & $\text{CG}_k(\lbrack 0, H_{\Omega} \rbrack)$ &
		$\text{DG}_{k-1}(\lbrack 0, H_{\Omega} \rbrack)$ &
		$\text{CG}_k(\triangle)$ & $\text{RT}_k(\triangle)$ & $\text{DG}_{k-1}(\triangle)$ \\
		$\text{BDFM}_{k+1}$ & $\text{CG}_{k+1}(\lbrack 0, H_{\Omega} \rbrack)$ &
		$\text{DG}_{k}(\lbrack 0, H_{\Omega} \rbrack)$ &
		$\text{CG}_{k+1}(\triangle)$ & $\text{BDFM}_{k+1}(\triangle)$ & $\text{DG}_{k}(\triangle)$ \\
		$\text{RTCF}_k$ & $\text{CG}_k(\lbrack 0, H_{\Omega} \rbrack)$ &
		$\text{DG}_{k-1}(\lbrack 0, H_{\Omega} \rbrack)$ &
		$\text{Q}_k(\square)$ & $\text{RTCF}_k(\square)$ & $\text{DQ}_{k-1}(\square)$
	\end{tabular}
\end{table}

\subsubsection{Compatible finite element discretization}\label{subsubsec:gw_disc}
We seek the unknown fields from \eqref{eq:gw_a}--\eqref{eq:gw_c} in the following finite element spaces:
\begin{equation}\label{eq:spaces}
	\vec{u} \in W_2^0,\quad
	p \in W_3,\quad
	b \in W_b,
\end{equation}
where $W_2^0$ is the subspace of $W_2$ satisfying the no-slip condition, and $W_b \equiv U_2\otimes V_0$.
Note that $W_b$ is just the vertical part of the velocity space
\footnote{The choice for $W_b$ in \eqref{eq:spaces} corresponds to a Charney-Phillips
	vertical staggering of the buoyancy variable, which is the desired approach for the UK Met Office's
	Unified Model \citep{melvin2010inherently}. One could also collocate $b$ with $p$
	($b \in W_3$), which corresponds to a Lorenz staggering. This however supports a
	computational mode which is exacerbated by fast-moving waves. We restrict our discussion
	to the former case.}.
That is, $W_b$ and $W_2^v$ have the same number of degrees of freedom, but differ in how
they are pulled back to the reference element. For ease of notation, we write $W_2$ in place
of $W_2^0$.

To obtain the discrete system, we simply multiply equations \eqref{eq:gw_a}--\eqref{eq:gw_c}
by test functions $\vec{w}\in W_2$, $\phi \in W_3$ and $\eta \in W_b$ and
integrate by parts. We introduce the increments $\delta \vec{u} \equiv \vec{u}^{(n+1)}-\vec{u}^{(n)}$,
and set $\vec{u}_0\equiv \vec{u}^{(n)}$ (similarly for $\delta p$, $p_0$, $\delta b$, and $b_0$).
Using an implicit midpoint rule discretization, we need to solve the following linear variational
problem at each time-step: find
$\delta\vec{u} \in W_2$, $\delta p \in W_3$ and $\delta b \in W_b$ such
that
\begin{align}
	\left(\vec{w},\delta\vec{u}\right)_{\mathcal{T}_h}
	+ \Delta t\left(\vec{w},\boldsymbol{\Omega}\times\delta\vec{u}\right)_{\mathcal{T}_h}
	- \frac{\Delta t}{2}\left(\nabla\cdot\vec{w},\delta p\right)_{\mathcal{T}_h}
	- \frac{\Delta t}{2}\left(\vec{w},\delta b\zhat\right)_{\mathcal{T}_h}
	&= r_u, \label{eq:inc-u} \\
	\left(\phi,\delta p\right)_{\mathcal{T}_h}
	+\frac{\Delta t}{2}c^2\left(\phi,\nabla\cdot\delta\vec{u}\right)_{\mathcal{T}_h}
	&= r_p, \label{eq:inc-p} \\
	\left(\eta,\delta b\right)_{\mathcal{T}_h}
	+ \frac{\Delta t }{2}N^2\left(\eta,\delta\vec{u}\cdot\zhat\right)_{\mathcal{T}_h}
	&= r_b. \label{eq:inc-b}
\end{align}
for all $\vec{w} \in W_2$, $\phi \in W_3$ and $\eta \in W_b$. The resulting matrix equations
have the form:
\begin{equation}\label{eq:mixed-gravity-wave}
	\begin{bmatrix}
	A_{\vec{u}} & -\frac{\Delta t}{2}D^T & -\frac{\Delta t}{2}Q^T \\
	\frac{\Delta t}{2}c^2D & M_p & 0 \\
	\frac{\Delta t}{2}N^2Q & 0 & M_b
	\end{bmatrix}
	\begin{Bmatrix}
	U\\ P\\ B
	\end{Bmatrix}
	=
	\begin{Bmatrix}
	R_{\vec{u}}\\ R_p\\ R_b
	\end{Bmatrix},
\end{equation}
where $A_{\vec{u}} = M_{\vec{u}} + \Delta t C_{\vec{\Omega}}$, $C_{\vec{\Omega}}$ is
the matrix associated with the Coriolis term, $M_{\vec{u}}$, $M_p$, $M_b$ are mass matrices,
$D$ is the weak divergence term, and $Q$ is the operator coupling $\delta \vec{u}$ and $\delta b$.

To solve \eqref{eq:mixed-gravity-wave}, we can use the buoyancy equation
to arrive at an elimination procedure by substituting:
\begin{equation}\label{eq:b-elim}
	\delta b = r_b - \frac{\Delta t}{2}N^2 \delta\vec{u}\cdot\zhat
\end{equation}
into \eqref{eq:inc-u}. Note that in a planar geometry, \eqref{eq:b-elim} is
satisfied point-wise. However, this is not true when orography is present or on the surface of a sphere. Hence this is a further approximation of the equations.
After solving for $\vec{u}$ abd $p$, we reconstruct $b$ from Equation
\eqref{eq:inc-b}
which requires solving the mass matrix $M_b$.
Fortunately, $M_b$ is well-conditioned (independent of the grid resolution/time-step) and decoupled columnwise; it
can be inverted with a small number of conjugate gradient iterations.

The full solution procedure is outlined as follows. First, we approximately eliminate
the buoyancy to obtain a mixed system for the velocity and pressure:
\begin{equation}\label{eq:mixed-vel-pr-system}
\mathcal{A}\begin{Bmatrix}
U \\ P
\end{Bmatrix} =
\begin{bmatrix}
\widetilde{A}_{\vec{u}} & -\frac{\Delta t}{2} D^T \\
c^2\frac{\Delta t}{2} D & M_p
\end{bmatrix}\begin{Bmatrix}
U \\ P
\end{Bmatrix} =
\begin{Bmatrix}
\widetilde{R}_{\vec{u}} \\ R_p
\end{Bmatrix},
\end{equation}
where $\widetilde{A}_{\vec{u}} = A_{\vec{u}} + \frac{\Delta t^2}{4} N^2 Q^T M_b^{-1} Q$
and $\widetilde{R}_{\vec{u}} = R_{\vec{u}} + \frac{\Delta t}{2} Q^T M_b^{-1} R_b$ is the
modified velocity operator and right-hand side respectively. Note that in our elimination strategy,
the expression for $\widetilde{A}_{\vec{u}}$ corresponds to the bilinear form obtained
after substituting the point-wise expression for $\delta b$:
\begin{equation}
	\widetilde{A}_{\vec{u}} \leftarrow
	\left(\vec{w},\delta\vec{u}\right)_{\mathcal{T}_h}
	+ \Delta t\left(\vec{w},\boldsymbol{\Omega}\times\delta\vec{u}\right)_{\mathcal{T}_h} +
	\frac{\Delta t^2}{4}N^2\left(\vec{w}\cdot\zhat, \delta\vec{u}\cdot\zhat\right)_{\mathcal{T}_h}.
\end{equation}
A similar construction holds for $\widetilde{R}_{\vec{u}}$. Once \eqref{eq:mixed-vel-pr-system} is
solved, $\delta b$ is reconstructed by solving:
\begin{equation}\label{eq:bsolver}
		M_b B + \frac{\Delta t}{2} N^2 Q U = R_b
		\quad\implies\quad
		B = M_b^{-1}R_b - \frac{\Delta t}{2} N^2 M_b^{-1} Q U.
\end{equation}
Equation \eqref{eq:bsolver} can be efficiently inverted using a preconditioned conjugate gradient
method (preconditioned by block ILU with zero in-fill).

\subsubsection{Preconditioning the mixed velocity pressure system}\label{subsubsec:gwprecon}
The primary difficulty is find robust solvers for \eqref{eq:mixed-vel-pr-system}. This
was studied by \cite{mitchell2016high} within the context of developing a robust
preconditioner to withstand fast-moving acoustic waves. This is critical, as
\eqref{eq:gw_a}--\eqref{eq:gw_c} support fast waves driven by perturbations to the
pressure field. However, the implicit treatment of the Coriolis term was not taken
into account. We consider two preconditioning strategies.

The first strategy follows from \cite{mitchell2016high}. As with the rotating shallow water
system in Section \ref{subsec:williamson5}, we can construct a preconditioner based on
the Schur-complement factorization of $\mathcal{A}$ in \eqref{eq:mixed-vel-pr-system}:
\begin{equation}\label{eq:inv-A-schur}
	\mathcal{A}^{-1} =
	\begin{bmatrix}
	I & \frac{\Delta t}{2}\widetilde{A}_{\vec{u}}^{-1}D^T \\
	0 & I
	\end{bmatrix}
	\begin{bmatrix}
	\widetilde{A}_{\vec{u}}^{-1} & 0 \\
	0 & H^{-1}
	\end{bmatrix}
	\begin{bmatrix}
	I & 0 \\
	-c^2\frac{\Delta t}{2}D\widetilde{A}_{\vec{u}}^{-1} & I
	\end{bmatrix},
\end{equation}
where $H = M_p + c^2\frac{\Delta t^2}{4}D\widetilde{A}_{\vec{u}}^{-1} D^T$ is
the dense pressure Helmholtz operator. Because we have chosen to include
the Coriolis term, the operator $H$ is non-symmetric, and we require
a sparse approximation to
\begin{equation}
	H = M_p + c^2\frac{\Delta t^2}{4}D\left(\widetilde{M}_{\vec{u}} + \Delta t C_{\vec{\Omega}}\right)^{-1} D^T
\end{equation}
where $\widetilde{M}_{\vec{u}}$ is the modified velocity mass matrix.
As $\Delta t$ increases, the contribution of $C_{\vec{\Omega}}$
becomes more prominent in $H$, making sparse approximations of $H$ more challenging.
We elaborate on this further below when we present the results of our second solver.

Our second strategy revolves around the hybridization of the system defined in
\eqref{eq:mixed-vel-pr-system}, using Firedrake's \texttt{HybridizationPC}. The space of traces is
generated on the faces of triangular-prisms and extruded quadrilaterals,
with appropriate polynomial degrees such that every trace function lies in
the same polynomial space as $\delta\vec{u}\cdot\vec{n}|_f$ for all faces $f$.

\begin{figure}[!htb]
\includegraphics[width=0.45\textwidth]{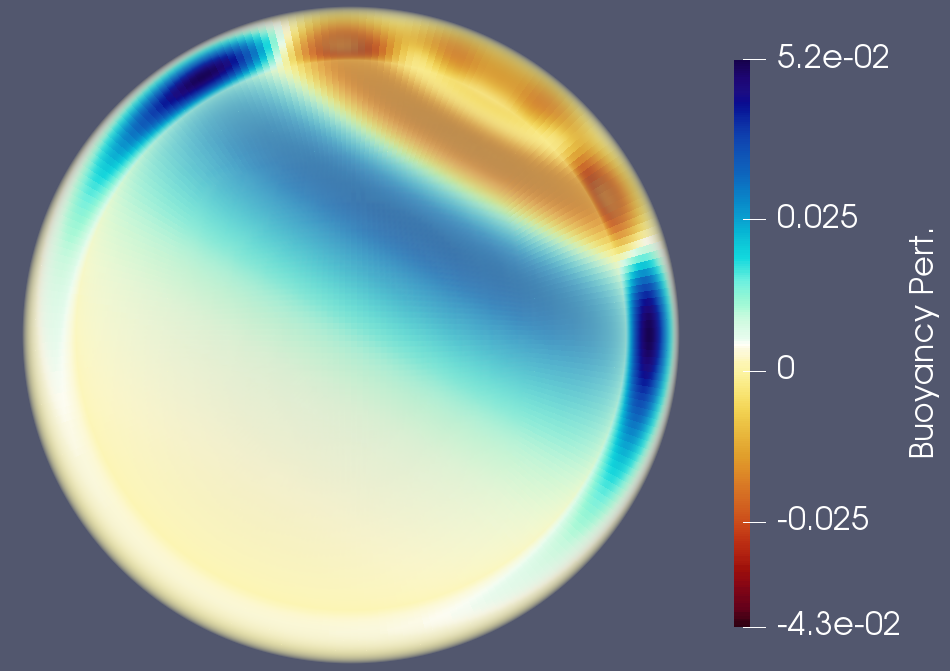}
\caption{Buoyancy perturbation ($y-z$ cross section) at $t=3600$s from a
	simple gravity wave test ($\Delta t = 100$s). The initial conditions (in lat.-long. coordinates)
	for the velocity is a simple solid-body rotation: $\vec{u} = 20\vec{e}_\lambda$, where $\vec{e}_\lambda$ is
	the unit vector pointing in the direction of decreasing longitude. A buoyancy anomaly is defined via
	$b = \frac{d^2}{d^2 + q^2}\sin(\pi z/10000)$, where $q = R\cos^{-1}(\cos(\phi)\cos(\lambda-\lambda_{\phi}))$,
	$d=5000$m, $R=6371\text{km}/125$ is the planet radius, and
	$\lambda_{\phi}=2/3$. The equations are discretized using the lowest-order method $\text{RTCF}_1$,
	with 24,576 quadrilateral cells in the horizontal and 64 extrusion levels.
	The velocity-pressure system is solved using hybridization.}
\label{fig:gwtest}
\end{figure}

We locally eliminate $U$ and $P$ after hybridization, and produce the resulting system for the traces:
\begin{equation}
	 H_\partial \Lambda = E, \quad H_\partial = K \widehat{\mathcal{A}}^{-1} K^T, \quad
	 E = K \widehat{\mathcal{A}}^{-1}
	\begin{Bmatrix}
	\widehat{R}_{\vec{u}} \\ R_p
	\end{Bmatrix},
\end{equation}
where $\widehat{\mathcal{A}}$ is the result of hybridizing $\mathcal{A}$ and $H_{\partial}$ is the statically
condensed trace operator to be inverted. The non-symmetric operator $H_{\partial}$ is inverted using a preconditioned
generalized conjugate residual (GCR) Krylov method, as suggested in \citet{thomas2003spectral}.
For our choice of preconditioner, we follow strategies outlined in \citet{elman2001multigrid}
and employ an algebraic multigrid method (V-cycle) with GMRES (five
iterations) smoothers on the coarse levels. The GMRES smoothers are preconditioned with block ILU on each level. For the finest level, block ILU
produces a line smoother (necessary for efficient solution on thin domains) when the trace variable nodes are numbered in vertical lines, as is the case in our Firedrake implementation. On the coarser levels,
less is known about the properties of ILU under the AMG coarsening strategies, but as we shall see, we observe performance that suggests ILU is still behaving
as a line smoother. More discussion
on multigrid for non-symmetric problems can be found in \citet{bramble1994uniform,bramble1988analysis,mandel1986multigrid}.
A gravity wave test using our solution strategy and hybridization preconditioner is illustrated in Figure \ref{fig:gwtest}
for a problem on a condensed Earth (radius scaled down by a factor of 125) and 10km lid.

\subsubsection{Robustness against acoustic Courant number with implicit Coriolis}
A desired property of the implicit linear solver is robustness with respect to the time-step size.
In particular, it is desirable for the execution time to remain constant across a wide range
of $\Delta t$. As $\Delta t$ increases, the
conditioning of the elliptic operator becomes worse. Therefore, iterative solvers need
to work much harder to reduce the residual down to a specified tolerance.

In this final experiment, we repeat a similar study to that presented in \citet{mitchell2016high}.
We fix the resolution of the problem and run our solver for a range of $\Delta t$. We measure
this by adjusting the horizontal acoustic Courant number $\lambda_C= c \frac{\Delta t}{\Delta x}$,
where $c$ is the speed of sound and $\Delta x$ is the horizontal resolution. We remark that
the range of Courant numbers used in this paper exceeds what is typical in operational forecast settings
(typically between $\mathcal{O}(2)$--$\mathcal{O}(10)$).
The grid set up mirrors that of actual global weather models; we extrude a spherical mesh of
the Earth upwards to a height of 85km. The set up for the different discretizations
(including degrees of freedom for the velocity-pressure and hybridized systems) is presented
in Table \ref{tab:3d-grids}.
\begin{table}[!htb]
	\captionsetup{position=top}
	\caption{Grid set up and discretizations for the acoustic Courant number study. The total
		unknowns (velocity and pressure) and hybridized unknowns (broken velocity, pressure,
		and trace) are shown in the last two columns (millions). The vertical resolution is fixed
		across all discretizations.}
	\label{tab:3d-grids}
	\centering
	\begin{tabular}{ccccccc}\hline
		\multicolumn{7}{c}{\textbf{Discretizations and grid information}} \\
		Mixed method & \# horiz. cells & \# vert. layers & $\Delta x$ & $\Delta z$ & $U$-$P$ dofs & Hybrid. dofs \\
		\hline
		$\text{RT}_1$ & 81,920 & 85 & 86 km & 1,000 m & 24.5 M & 59.3 M \\
		$\text{RT}_2$ & 5,120 & 85 & 346 km & 1,000 m & 9.6 M & 17.4 M \\
		$\text{BDFM}_2$ & 5,120 & 85 & 346 km & 1,000 m & 10.5 M & 18.3 M \\
		$\text{RTCF}_1$ & 98,304 & 85 & 78 km & 1,000 m & 33.5 M & 83.7 M \\
		$\text{RTCF}_2$ & 6,144 & 85 & 312 km & 1,000 m & 16.7 M & 29.3 M \\
		\hline
	\end{tabular}
\end{table}

\begin{figure}[!htbp]
	\centering
	\includegraphics[width=0.45\textwidth]{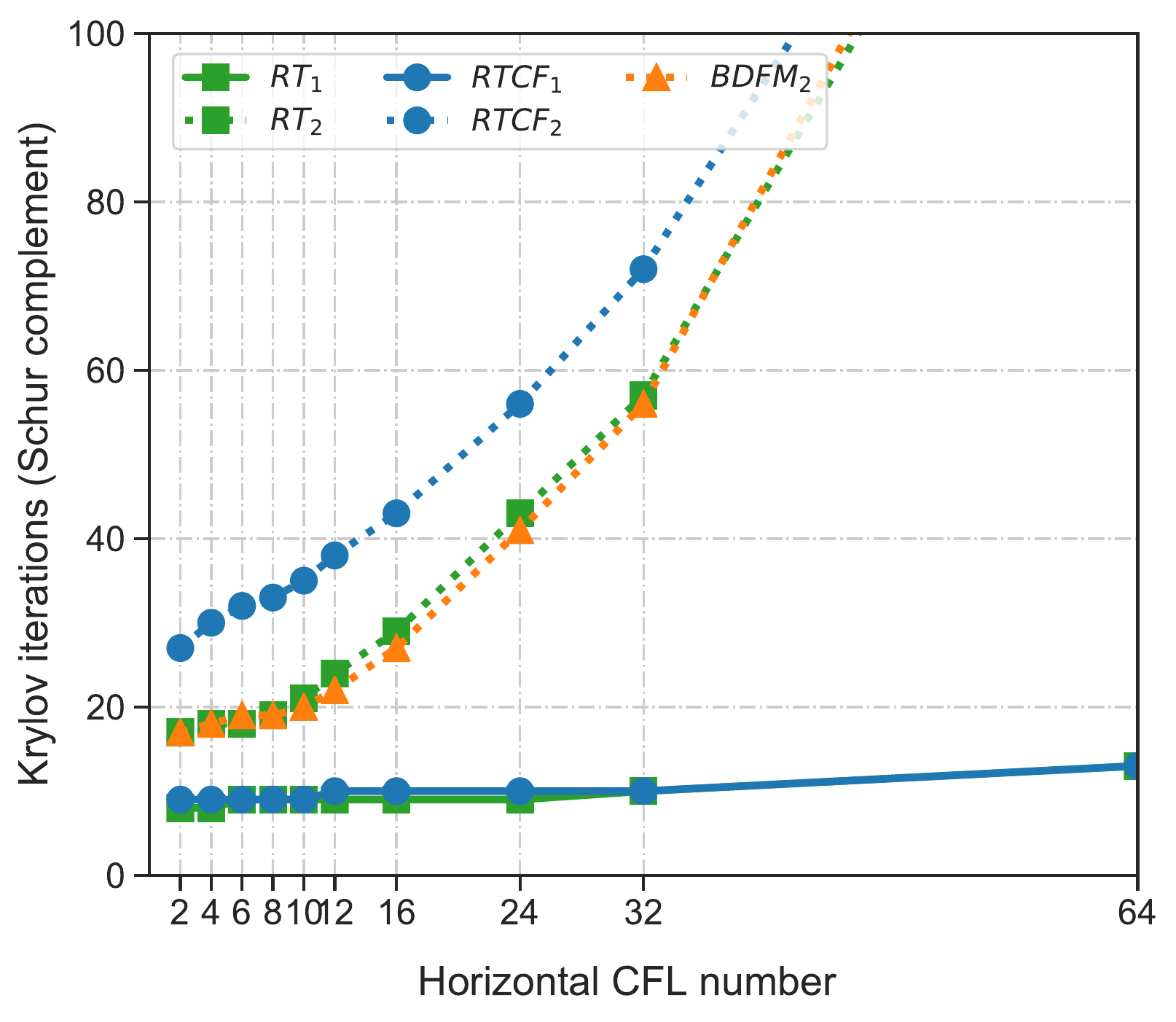}
	\caption{Number of Krylov iterations to invert the Helmholtz system using $\widetilde{H}^{-1}$
		as a preconditioner. The preconditioner is applied using a direct LU factorization within
		a GMRES method on the entire pressure equation. While the lowest-order methods grow slowly over
		the Courant number range, the higher-order (by only one approximation order)
		methods quickly degrade and diverge after the critical range
		$\lambda_C = \mathcal{O}(2)$--$\mathcal{O}(10)$. At $\lambda_C>32$, the solvers take over 150 iterations.}
	\label{fig:approx-schur-gw}
\end{figure}

It was shown in \citet{mitchell2016high} that using a sparse approximation of the
pressure Schur-complement of the form:
\begin{equation}\label{eq:diagH}
\widetilde{H} = M_p + c^2\frac{\Delta t^2}{4}D\left(\text{Diag}(\widetilde{M}_{\vec{u}})\right)^{-1} D^T
\end{equation}
served as a good preconditioner, leading to a system that was amenable to multigrid
methods and resulted in a Courant number independent solver. However, when the Coriolis term is included,
this is no longer the case: the
diagonal approximation to $\widetilde{M}_{\vec{u}}$ becomes worse with
increasing $\lambda_C$. To demonstrate this, we solve the mixed
problem on a low-resolution grid (10km lid, 10 vertical levels,
maintaining the same cell aspect ratio as in Table
\ref{tab:3d-grids}) using the Schur-complement factorization
\eqref{eq:inv-A-schur}, and
use LU factorizations to apply both $\widetilde{A}_{\vec{u}}^{-1}$ and
$\widetilde{H}^{-1}$. $H^{-1}$ is computed using preconditioned GMRES
iterations, and a flexible-GMRES algorithm is used on the full
velocity-pressure system.
If $\widetilde{H}^{-1}$ is a good approximation
to $H^{-1}$ we should see low iteration counts when inverting $H$.
Figure \ref{fig:approx-schur-gw} shows the results of this study for a
range of Courant numbers.

For the lower-order methods, the number of iterations to invert $H$
grow slowly but remain under control. Increasing the approximation degree
by one results in degraded performance. As $\Delta t$ increases, the number of Krylov iterations
needed to invert the system to a relative tolerance of $10^{-5}$ grows rapidly. It is clear that
this sparse approximation is not robust against Courant number. This can be explained by the fact that diagonalizing
the velocity operator fails to take into account the effects of the Coriolis term (which appear in off-diagonal
positions in the operator). Even if one were to use traditional mass-lumping (row-wise sums), the Coriolis
effects are effectively cancelled out due to asymmetry.

\begin{figure}[!htbp]
	\centering
	\begin{subfigure}{.5\textwidth}
		\centering
		\captionsetup{width=\linewidth}
		\includegraphics[width=0.85\textwidth]{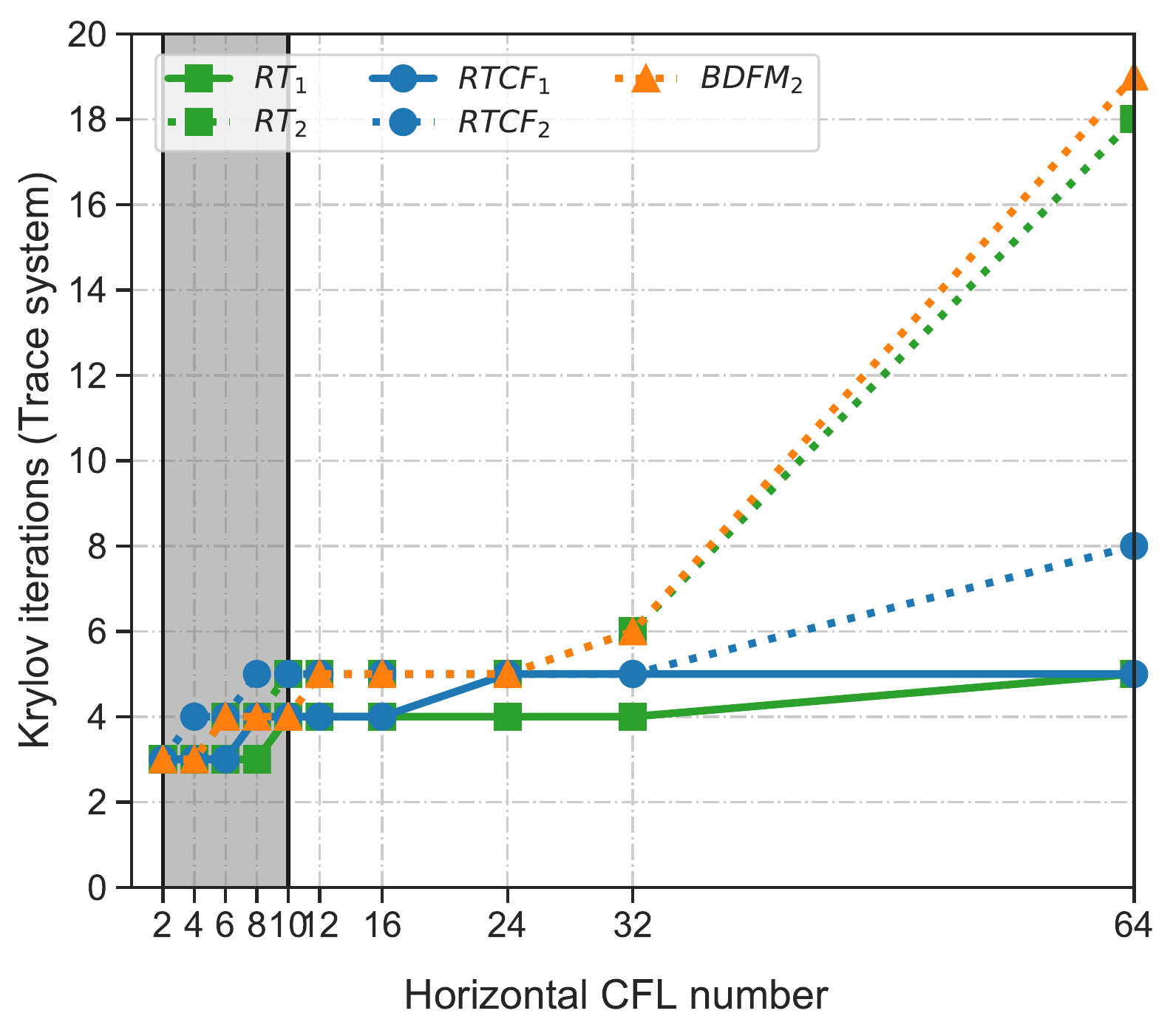}
		\caption{GCR iterations vs Courant number.}
		\label{fig:hybrid_itervscfl}
	\end{subfigure}%
	\begin{subfigure}{.5\textwidth}
		\centering
		\captionsetup{width=\linewidth}
		\includegraphics[width=0.85\textwidth]{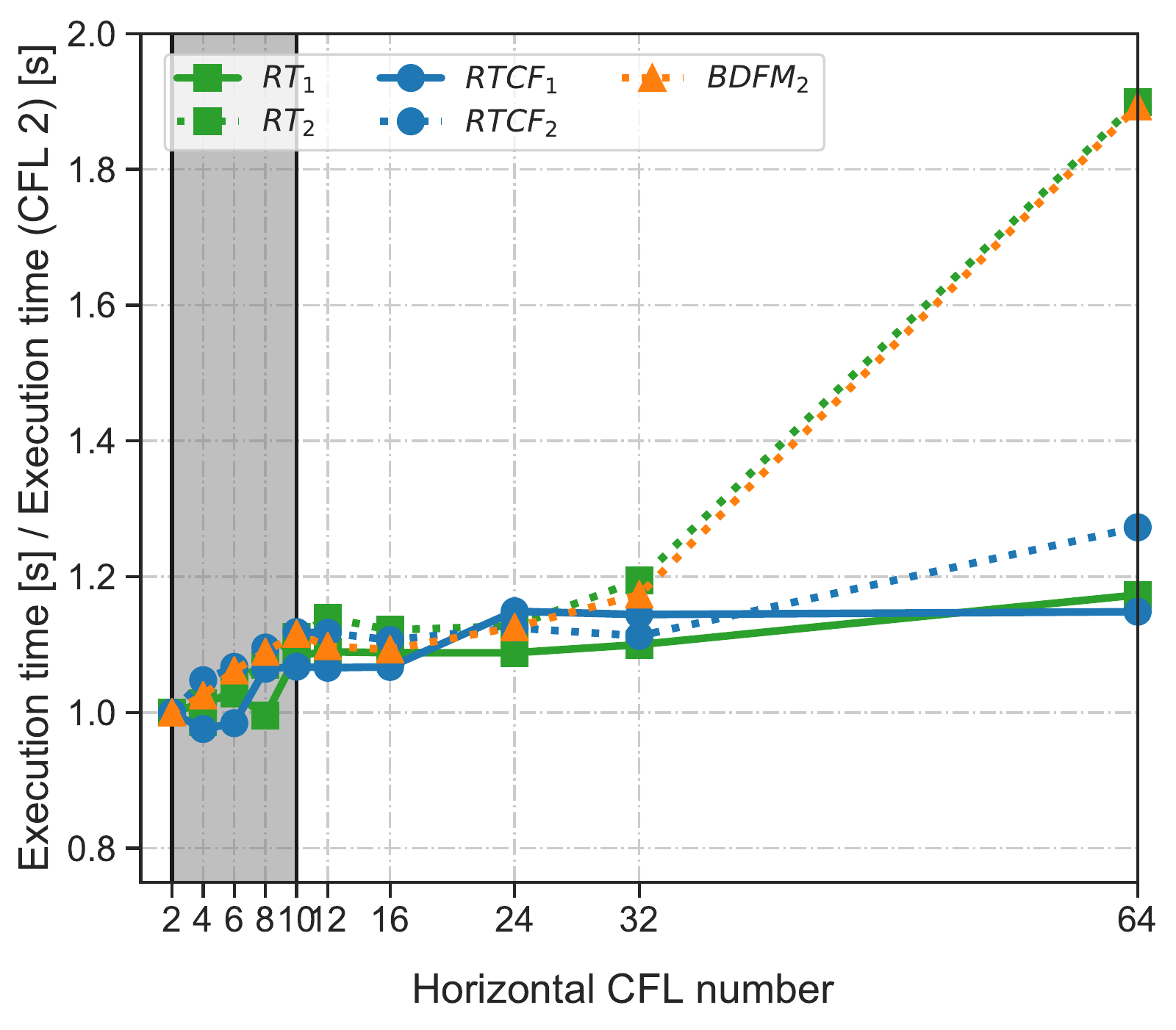}
		\caption{Work vs Courant number.}
		\label{fig:hybrid_rel_work}
	\end{subfigure}
	\caption{Courant number parameter test run on a fully-loaded compute node. Both figures display the hybridized
		solver for each discretization, described in Table \ref{tab:3d-grids}. The left figure (a)
		displays to total iteration count (preconditioned GCR) to solve the trace system to a
		relative tolerance of $10^{-5}$. The right figure (b) displays the relative work of each solver.
		Figure (b) takes into account not just the time-to-solution of the trace solver, but also the time
		required to forward eliminate and locally recover the velocity and pressure.
		}
	\label{fig:hybridwork}
\end{figure}

Hybridization avoids this problem: we always construct an
exact Schur complement, and only have to worry about solving the trace
system. We now show that this approach (described in
Section \ref{subsubsec:gwprecon}) is much more robust to
increases in the Courant number. We use
the same workstation as for the three-dimensional CG/HDG problem in Section \ref{subsec:HDGCGcomp}
(executed with a total of 40 MPI processes). Figure
\ref{fig:hybridwork} shows the parameter test for all the discretizations described in Table \ref{tab:3d-grids}.
We see that, in terms of total number of GCR iterations needed to invert the trace system, hybridization
is far more robust against Courant number than the approximate Schur
complement approach. They largely remain constant throughout the entire parameter range,
only varying by an iteration or two. It is not until after $\lambda_C >32$ that
we begin to see a larger jump
in the number of GCR iterations. This is expected, since the Coriolis operator
causes the problem to become singular for very large Courant numbers.
However, unlike with the approximate Schur-complement solver,
iteration counts are still
under control. In particular, each method (lowest and higher order) remains constant throughout the
critical range (shaded in gray in Figures \ref{fig:hybrid_itervscfl} and \ref{fig:hybrid_rel_work}).

In Figure \ref{fig:hybrid_rel_work}, we display the ratio of execution time and the time-to-solution
at the lowest Courant number of two. We perform this normalization to better compare the lower and higher
order methods (and discretizations on triangular prisms vs extruded quadrilaterals). The calculation
of the ratios include the time needed to eliminate and reconstruct the hybridized velocity/pressure variables.
The fact that the hybridization solver remains close to one demonstrates that the entire
solution procedure is largely $\lambda_C$-independent until around $\lambda_C=32$.
The overall trend is largely the same as what is observed in Figure \ref{fig:hybrid_itervscfl}. This
is due to our hybridization approach being solver dominated, with local operations like forward elimination
together with local recovery taking approximately $1/3$ of the total execution time for each method.
The percentage breakdown of the hybridization solver is similar to what is already presented in
Section \ref{subsubsec:hdgbreakdown}.

Implicitly treating the Coriolis term has been discussed for semi-implicit discretizations
of large-scale geophysical flows \citep{temperton1997treatment,cote1988two,cullen2001alternative,nechaev2004approximation}.
Incorporating the Coriolis term in finite element discretizations is difficult, as this makes the
challenge of find robust sparse approximation of the resulting elliptic operator even
more difficult. Hybridization shows promise here, as we no longer require the inversion
of a dense elliptic system. Instead, hybridization allows for the assembly of an elliptic
equation that both captures the effects of rotation and results in a sparse linear system.
In fact, the hybridization process mimics standard staggered finite different elimination
procedures used in many global circulation models. In other words, hybridization is the
finite element analogue of point-wise elimination strategies in finite difference codes.

\conclusions  %% \conclusions[modified heading if necessary]
\label{sec:conclusions}
We have presented Slate, and shown how this language can be used
to create concise mathematical representations of localized linear algebra on the tensors
corresponding to finite element forms. We have shown how this DSL can be
used in tandem with UFL in Firedrake to implement
solution approaches making use of automated code generation for
static condensation, hybridization, and localized post-processing. Setup and configuration
is done at runtime, allowing one to switch in different discretizations at will. In particular,
this framework alleviates much of the difficulty in implementing such methods
within intricate numerical, and paves the way for future low-level optimizations.
In this way, the framework in this paper can be used to help enable the rapid
development and exploration of new hybridization and static condensation techniques
for a wide class of problems.
We remark here that the reduction of global matrices via element-wise algebraic static
condensation, as described in \citet{guyan, irons} is also possible using Slate,
including other more general static condensation procedures outside the context of hybridization.

Our approach to preconditioner design revolves around its composable nature, in that
these Slate-based implementations can be seamlessly incorporated into complicated solution
schemes. In particular, there is current research in the design of dynamical cores for
numerical weather prediction using implementations of hybridization and static condensation
with Slate \citep{bauer2018energy,shipton2018higher}. The performance of such methods for
geophysical flows are a subject of on-going investigation.

In this paper, we have provided some examples of hybridization procedures for
compatible finite element discretizations of geophysical flows. These approaches
avoid the difficulty in constructing sparse approximations of dense elliptic operators.
Static condensation arising from hybridizable formulations can best be interpreted as
producing an \emph{exact} Schur-complement factorization on the global hybridizable system.
This eliminates the need for outer iterations from a suitable Krylov method to solve
the full mixed system, and replaces the original global mixed equations
with a condensed elliptic system. More extensive performance benchmarks, which requires
detailed analysis of the resulting operator systems arising from hybridization,
is a necessary next-step to determine whether hybridization provides
a scalable solution strategy for compatible finite elements in operational settings.

%% The following commands are for the statements about the availability of data sets and/or software code corresponding to the manuscript.
%% It is strongly recommended to make use of these sections in case data sets and/or software code have been part of your research the article is based on.

\codedataavailability{The contribution in this paper is available through open-source software
	provided by the Firedrake Project: \url{https://www.firedrakeproject.org/}.
	We cite archives of the exact software versions used to produce the results in
	this paper. For all components of the Firedrake project used in this paper, see
	\cite{zenodo-meta}. The numerical experiments, full solver configurations,
	code-verification (including local-processing),
	and raw data are available in \cite{zen-tabularasa}.}

\appendix

\section{Semi-implicit method for the shallow water system}\label{app:semiimplicit}
For some tessellation, $\mathcal{T}_h$, our semi-discrete mixed method for
\eqref{eq:nswe-a}--\eqref{eq:nswe-b} seeks approximations
$(\boldsymbol{u}_h, D_h) \in \boldsymbol{U}_h \times V_h \subset
\boldsymbol{H}(\text{div})\times L^2$ satisfying:
\begin{align}
	\left(\boldsymbol{w}, \frac{\partial \boldsymbol{u}_h}{\partial t}\right)_{\mathcal{T}_h} -
	\left(\nabla^\perp\left(\boldsymbol{w}\cdot \boldsymbol{u}_h^\perp\right), \boldsymbol{u}_h^\perp\right)_{\mathcal{T}_h} + 
	\left(\boldsymbol{w}, f\boldsymbol{u}_h^\perp\right)_{\mathcal{T}_h}
	+ \facetinner{\jump{\boldsymbol{n}^\perp
			\boldsymbol{w}\cdot \boldsymbol{u}_h^\perp
		},
		\tilde{\boldsymbol{u}}_h^{\perp}}_{\partial\mathcal{T}_h}
	\nonumber        \\
	-\left(\nabla\cdot\boldsymbol{w}, g\left(D_h + b\right) + \frac{1}{2}|\boldsymbol{u}_h|^2\right)_{\mathcal{T}_h}
	&= 0, &\forall \boldsymbol{w} \in \boldsymbol{U}_h,
	\label{eq:mmnswe-a}\\
	\left(\phi, \frac{\partial D_h}{\partial t}\right)_{\mathcal{T}_h}
	- \cellinner{\nabla\phi, \boldsymbol{u}_h D_h}_{\mathcal{T}_h}
	+ \facetinner{\jump{\phi
			\boldsymbol{u}_h},\tilde{D}_h}_{\partial\mathcal{T}_h},
	&= 0, &\forall \phi \in V_h,
	\label{eq:mmnswe-b}
\end{align}
where $\tilde{\cdot}$ indicates that the value of the function
should be taken from the upwind side of each facet. The discretisation
of the velocity advection operator is an extension of the
energy-conserving scheme of \citet{natale2017variational} to the
shallow-water equations.

The time-stepping scheme follows a Picard iteration semi-implicit approach, where
predictive values of the relevant fields are determined via an explicit step of the
advection equations, and corrective updates are generated by solving
an implicit linear system (linearized about a state of rest) for
$(\Delta\boldsymbol{u}_h,\Delta D_h)\in \boldsymbol{U}_h \times V_h$,
given by
\begin{align}
	\left(\boldsymbol{w}, \Delta\boldsymbol{u}_h\right)_{\mathcal{T}_h}
	+ \frac{\Delta t}{2}\left(\boldsymbol{w},
	f\Delta\boldsymbol{u}_h^\perp \right)_{\mathcal{T}_h}
	- \frac{\Delta t}{2}\left(\nabla\cdot\boldsymbol{w}, g\Delta D_h
	\right)_{\mathcal{T}_h}
	&= -R_{\boldsymbol{u}}[\boldsymbol{u}_h^{n+1},D_h^{n+1};\boldsymbol{w}],
	&\forall \boldsymbol{w} \in \boldsymbol{U}_h,
	\label{eq:nswe-weak-a}\\
	\cellinner{\phi, \Delta D_h}_{\mathcal{T}_h}
	+ \frac{H\Delta t}{2}\cellinner{\phi, \nabla\cdot\Delta\boldsymbol{u}_h}_{\mathcal{T}_h} &= -R_{D}[\boldsymbol{u}_h^{n+1},D_h^{n+1};\phi],
	&\forall \phi \in V_h,
	\label{eq:nswe-weak-b}
\end{align}
where $H$ is the mean layer depth, and $R_{\boldsymbol{u}}$ and
$R_D$ are residual linear forms that vanish when
$\boldsymbol{u}_h^{n+1}$ and $D_h^{n+1}$ are solutions to the implicit
midpoint rule time discretization of \eqref{eq:mmnswe-a}--\eqref{eq:mmnswe-b}.
The residuals are evaluated using the predictive values of $\boldsymbol{u}_h^{n+1}$
and $D_h^{n+1}$.

The implicit midpoint rule time discretization of the non-linear
rotating shallow water equations \eqref{eq:mmnswe-a}--\eqref{eq:mmnswe-b} is:
\begin{align}
	\left(\boldsymbol{w}, \boldsymbol{u}^{n+1}_h-\boldsymbol{u}_h^n\right)_{\mathcal{T}_h} -
	\Delta t\left(\nabla^\perp\left(\boldsymbol{w}\cdot {\boldsymbol{u}_h^*}^\perp\right), {\boldsymbol{u}_h^*}^\perp\right)_{\mathcal{T}_h} + 
	\Delta t\left(\boldsymbol{w}, f{\boldsymbol{u}_h^*}^\perp\right)_{\mathcal{T}_h}
	\nonumber \\
	+ \Delta t\facetinner{\jump{\boldsymbol{n}^\perp
			\boldsymbol{w}\cdot {\boldsymbol{u}_h^*}^\perp
		},
		{\boldsymbol{\tilde{u}}_h^*}^\perp}_{\partial\mathcal{T}_h} \nonumber \\
	-\Delta t\left(\nabla\cdot\boldsymbol{w}, g\left(D_h^* + b\right) +
	\frac{1}{2}|\boldsymbol{u}_h^*|^2\right)_{\mathcal{T}_h}
	&= 0, &\forall \boldsymbol{w} \in \boldsymbol{U}_h,
	\label{eq:mmnswe-a-dt}\\
	\left(\phi, D_h^{n+1}-D_h^n\right)_{\mathcal{T}_h}
	- \Delta t
	\cellinner{\nabla\phi, \boldsymbol{u}_h^* D_h^*}_{\mathcal{T}_h}
	+ \Delta t\facetinner{\jump{\phi
			\boldsymbol{u}_h^*},{\tilde{D}}_h^*}_{\partial\mathcal{T}_h},
	&= 0, &\forall \phi \in V_h,
	\label{eq:mmnswe-b-dt}
\end{align}
where $\boldsymbol{u}_h^* = (\boldsymbol{u}_h^{n+1}+\boldsymbol{u}_h^n)/2$
and ${D}_h^* = ({D}_h^{n+1}+{D}_h^n)/2$.

One approach to construct the residual functionals $R_{\boldsymbol{u}}$ and
$R_D$ would be to simply define these from
\eqref{eq:mmnswe-a-dt}--\eqref{eq:mmnswe-b-dt}. However, this leads to a small
critical time-step for stability of the scheme. To make the numerical scheme more stable,
we define residuals as follows. For $R_{\boldsymbol{u}}$, we first solve
for $\boldsymbol{v}_h \in \boldsymbol{U}_h$ such that
\begin{align}
	\left(\boldsymbol{w}, \boldsymbol{v}_h-\boldsymbol{u}_h^n\right)_{\mathcal{T}_h} -
	\Delta t\left(\nabla^\perp\left(\boldsymbol{w}\cdot {\boldsymbol{u}_h^*}^\perp\right), {\boldsymbol{v}_h^\sharp}^\perp\right)_{\mathcal{T}_h} + 
	\Delta t\left(\boldsymbol{w}, f{\boldsymbol{v}_h^\sharp}^\perp\right)_{\mathcal{T}_h} \nonumber\\
	+ \Delta t\facetinner{\jump{\boldsymbol{n}^\perp
			\boldsymbol{w}\cdot {\boldsymbol{u}_h^*}^\perp
		},
		{\boldsymbol{\tilde{v}}^{\sharp}_h}^\perp}_{\partial\mathcal{T}_h}\nonumber\\
	-\Delta t\left(\nabla\cdot\boldsymbol{w}, g\left(D_h^* + b\right) + \frac{1}{2}|\boldsymbol{u}_h^*|^2\right)_{\mathcal{T}_h}
	&= 0, \quad \forall \boldsymbol{w} \in \boldsymbol{U}_h, \label{eq:res-lvp-v}
\end{align}
where $\boldsymbol{v}_h^\sharp=(\boldsymbol{v}_h + \boldsymbol{u}_h^n)/2$.
This is a linear variational problem. Then,
\begin{equation}
	R_{\boldsymbol{u}}[\boldsymbol{u}_h^{n+1},D_h^{n+1};\boldsymbol{w}]
	= \left(\boldsymbol{w}, \boldsymbol{v}_h - \boldsymbol{u}_h^{n+1}\right)
	_{\mathcal{T}_h}.
\end{equation}
Similarly, for $R_{D}$ we first solve for $E_h\in V_h$ such that
\begin{equation}\label{eq:res-lvp-E}
	\left(\phi, E_h-D_h^n\right)_{\mathcal{T}_h}
	- \Delta t
	\cellinner{\nabla\phi, \boldsymbol{u}_h^* E_h^\sharp}_{\mathcal{T}_h}
	+ \Delta t\facetinner{\jump{\phi
			\boldsymbol{u}_h^*},{\tilde{E}}_h^\sharp}_{\partial\mathcal{T}_h} = 0,
	\quad\forall \phi \in V_h,
\end{equation}
where $E_h^\sharp=(E_h + D_h^n)/2$. This is also a linear problem. Then, 
\begin{equation}
	R_{D}[\boldsymbol{u}_h^{n+1},D_h^{n+1};\phi]
	= \left(\phi, E_h - D_h^{n+1}\right)
	_{\mathcal{T}_h}.
\end{equation}

This  process can be thought of as iteratively solving for the average velocity and depth
that satisfies the implicit midpoint rule discretisation. Both \eqref{eq:res-lvp-v} and \eqref{eq:res-lvp-E}
can be solved separately, since there is no coupling between them.
The fields $\boldsymbol{v}_h$ and $E_h$ are then used to
construct the right-hand side for the implicit linearized system in
\eqref{eq:nswe-weak-a}--\eqref{eq:nswe-weak-b}. Once the system is solved, the
solution $(\Delta\boldsymbol{u}_h,\Delta D_h)$ is then used to update
the iterative values of $\boldsymbol{u}_h^{n+1}$ and $D_h^{n+1}$
according to $(\boldsymbol{u}_h^{n+1},D_h^{n+1})\mapsfrom
(\boldsymbol{u}_h^{n+1} + \Delta \boldsymbol{u}_h,D_h^{n+1} + \Delta
D_h)$, having initially chosen
$(\boldsymbol{u}_h^{n+1},D_h^{n+1})=(\boldsymbol{u}_h^{n},D_h^{n})$.

\noappendix       %% use this to mark the end of the appendix section

%% Regarding figures and tables in appendices, the following two options are possible depending on your general handling of figures and tables in the manuscript environment:

%% Option 1: If you sorted all figures and tables into the sections of the text, please also sort the appendix figures and appendix tables into the respective appendix sections.
%% They will be correctly named automatically.

%% Option 2: If you put all figures after the reference list, please insert appendix tables and figures after the normal tables and figures.
%% To rename them correctly to A1, A2, etc., please add the following commands in front of them:

\appendixfigures  %% needs to be added in front of appendix figures

\appendixtables   %% needs to be added in front of appendix tables

%% Please add \clearpage between each table and/or figure. Further guidelines on figures and tables can be found below.

\authorcontribution{T. H. Gibson is the principal author and developer of the software presented in
this paper and main author of the text. Authors L. Mitchell and D. A. Ham assisted and guided the software abstraction
as a domain-specific language, and edited text. C. J. Cotter contributed to
the formulation of the geophysical fluid dynamics and the design of the
numerical experiments, and edited text.}

\competinginterests{D. A. Ham is an executive editor of the journal. The other authors declare they
have no other competing interests.}

%\disclaimer{Wash hands before and after use.} %% optional section

\begin{acknowledgements}
This work was supported by the Engineering and Physical Sciences Research Council
(grant numbers: EP/M011054/1, EP/L000407/1, and EP/L016613/1), and the
Natural Environment Research Council (grant number: NE/K008951/1).
\end{acknowledgements}

%% REFERENCES

%% The reference list is compiled as follows:
%\begin{thebibliography}{}
%
%\bibitem[AUTHOR(YEAR)]{LABEL1}
%REFERENCE 1
%
%\bibitem[AUTHOR(YEAR)]{LABEL2}
%REFERENCE 2
%
%\end{thebibliography}

%% Since the Copernicus LaTeX package includes the BibTeX style file copernicus.bst,
%% authors experienced with BibTeX only have to include the following two lines:
%%
\bibliographystyle{copernicus}
\bibliography{bibliography.bib}

\end{document}